\documentclass[sigconf,10pt]{acmart}

\usepackage[OT1]{fontenc}
\usepackage{algorithm}
\usepackage{algorithmicx}
\usepackage[noend]{algpseudocode}
\usepackage{textcomp}
\usepackage{booktabs} 
\usepackage{comment}
\usepackage{wrapfig}
\usepackage{xcolor}
\usepackage{graphicx}
\usepackage{subfig}
\usepackage{array}
\usepackage{multicol}
\usepackage{afterpage}
\usepackage{float}
\usepackage{listings}
\usepackage{multirow}
\usepackage{arydshln}
\usepackage{xspace}
\usepackage{enumitem}
\usepackage{fancyhdr}
\usepackage{amsmath}
\usepackage{url}
\usepackage[htt]{hyphenat}

\usepackage{cleveref}
\crefformat{chapter}{\S#2#1#3}
\crefformat{section}{\S#2#1#3} 
\crefformat{subsection}{\S#2#1#3}
\crefformat{subsubsection}{\S#2#1#3}

\graphicspath{{figures/}}

\setcopyright{rightsretained}

\copyrightyear{2020} 
\acmYear{2020} 
\setcopyright{acmcopyright}\acmConference[SoCC '20]{ACM Symposium on Cloud Computing}{October 19--21, 2020}{Virtual Event, USA}
\acmBooktitle{ACM Symposium on Cloud Computing (SoCC '20), October 19--21, 2020, Virtual Event, USA}
\acmPrice{15.00}
\acmDOI{10.1145/3419111.3421299}
\acmISBN{978-1-4503-8137-6/20/10}

\begin{document}

\title{Semi-Dynamic Load Balancing: Efficient Distributed Learning in Non-Dedicated Environments}

\author{Chen Chen}
\affiliation{%
  \institution{HKUST}
}
\email{cchenam@connect.ust.hk}

\author{Qizhen Weng}
\affiliation{
	\institution{HKUST}
}
\email{qwengaa@cse.ust.hk}

\author{Wei Wang}
\affiliation{
	\institution{HKUST}
}
\email{weiwa@cse.ust.hk}

\author{Baochun Li}
\affiliation{
	\institution{University of Toronto}
}
\email{bli@ece.toronto.edu}

\author{Bo Li}
\affiliation{
	\institution{HKUST}
}
\email{bli@cse.ust.hk}

\begin{abstract}
Machine learning (ML) models are increasingly trained in clusters with non-dedicated workers possessing heterogeneous resources. In such scenarios, model training efficiency can be negatively affected by \emph{stragglers}---workers that run much slower than others. Efficient model training requires eliminating such stragglers, yet for modern ML workloads, existing load balancing strategies are inefficient and even infeasible. In this paper, we propose a novel strategy called \emph{semi-dynamic load balancing} to eliminate stragglers of distributed ML workloads. The key insight is that ML workers shall be load-balanced at \emph{iteration boundaries}, being non-intrusive to intra-iteration execution. We develop LB-BSP based on such an insight, which is an integrated worker coordination mechanism that adapts workers' load to their instantaneous processing capabilities by right-sizing the sample batches at the synchronization barriers. We have custom-designed the batch sizing algorithm respectively for CPU and GPU clusters based on their own characteristics. LB-BSP has been implemented as a Python module for ML frameworks like TensorFlow and PyTorch. Our EC2 deployment confirms that LB-BSP is practical, effective and light-weight, and is able to accelerating distributed training by up to $54\%$.
\end{abstract}



\maketitle

\section{Introduction}
\label{sec:intro}

Machine learning (ML) models, such as deep neural networks, are widely used for a range of applications to attain state-of-the-art performance~\cite{Krizhevsky12a,Farabet13a,Collobert11a,Sutskever14a}. Owing to their compute-intensive nature, ML models are often trained in a \emph{distributed} manner~\cite{Dean12a,Li14a,Chilimbi14a,abadi2016tensorflow}: many \emph{worker} threads iteratively process small subsets of the training data (i.e., \emph{sample batches}), and use the computed updates to refine the global model parameters.

With the surge of ML training demands, it has become increasingly common to serve ML jobs in \emph{non-dedicated} environments, e.g., with {\em heterogeneous} hardware from spot markets~\cite{gpu_spot}, or with \emph{time-varying} resources from shared production clusters~\cite{jeon2018multi,xiao2018gandiva,hazelwood2018applied}. In such cases, workers with less capable resources would progress slower and become \emph{stragglers}. Under the popular \emph{Bulk Synchronous Parallel} (BSP) scheme, fast workers have to wait for slow ones at the end of each iteration, and stragglers would thus impair the model training efficiency.

To mitigate the negative effect of stragglers, a large number of mechanisms have been developed in the literature. For example, some~\cite{Dean12a,ho2013more,Cui14a} have proposed to relax the synchronization barriers to avoid end-of-iteration resource wastage, yet this negatively affects the update quality and requires more iterations for models to converge. In fact, stragglers in non-dedicated clusters are mainly caused by the \emph{mismatch} between workers' load and their processing capability. In response, the most effective strategy to eliminate stragglers is to balance the load of workers according to their instantaneous processing capability.

Nonetheless, ML workloads pose new challenges to the load balancing community. Existing load balancing schemes designed for traditional multi-core or big data scenarios can be broadly classified into two categories: \emph{static} and \emph{dynamic} load balancing~(\cref{sec:load-balancing}). Yet, different from those traditional workloads such as HPC processing~\cite{blumofe1999scheduling,dinan2009scalable} or MapReduce~\cite{dean2010mapreduce}, ML computations are highly \emph{structured} (with thousands of short iterations) and also \emph{tensor-based} (samples in each iteration are packaged into a \emph{non-divisible} matrix for fast processing). Static load balancing approaches~\cite{kokilavani2011load,samal2013analysis,tang2000optimizing} are not aware of runtime resource variations, and dynamic approaches~\cite{yang2018scheduling,acun2016mitigating,harlap2016addressing}, which are mainly based on \emph{work stealing}~\cite{blumofe1999scheduling,dinan2009scalable,acar2013scheduling}, are also deficient for ML workloads. First, work stealing usually requires fine-grained worker progress monitoring and runtime load migration, which is inefficient for an iterative model training process. Second, runtime load migration assumes that samples are processed \emph{one by one}, which is indeed incompatible with the style of \emph{tensor-based} processing in modern ML frameworks~\cite{abadi2016tensorflow,chen2015mxnet,PyTorch}.

In this paper, we design a new load balancing strategy for distributed model training workloads, called \emph{semi-dynamic load balancing}. In broad strokes, the high-level idea is to perform all load balancing actions---worker monitoring, straggler identification and load redistribution---on the iteration boundaries, with load adjustment enforced by tuning each worker's sample \emph{batch size}. Following such a high-level idea, we propose \emph{Load-Balanced Bulk Synchronous Parallel} (LB-BSP), a composite scheme atop BSP that seeks to equalize all the workers' batch processing times by speculatively configuring their batch sizes at the synchronization barriers. An immediate question is then how to set the batch size at the synchronization barriers for the best load balancing effect in the upcoming iteration. This evolves into different challenges for CPU and GPU clusters that are two typical platform types for distributed learning.

For a CPU worker, our profiling work~(\cref{sec:profile_cpu}) shows that its batch processing time is proportional to its batch size, with the proportion coefficient representing the instantaneous sample processing capability. In shared clusters, such a coefficient may vary drastically with the temporal resources~\cite{hazelwood2018applied,reiss2012heterogeneity}, and thus shall be predicted prior to each iteration. For this purpose, we employ a special kind of recurrent neural network called NARX~\cite{diaconescu2008use}, which can make accurate performance predictions by taking into account the driving resources such as CPU and memory.

In multi-tenant GPU clusters~\cite{jeon2018multi,peng2018optimus,xiao2018gandiva,gu2019tiresias}, a GPU is usually dedicated to one worker without sharing. But the relationship between a GPU worker's batch processing time and batch size is not proportional and difficult to profile at runtime~(\cref{sec:profile_gpu}). Therefore, instead of the profiling-based analytical methods, we propose an \emph{iterative} batch size tuning algorithm that can efficiently approximate the load-balanced state without prior knowledge.

Furthermore, while our method of worker-adaptive batch sizing can effectively eliminate stragglers, it inevitably results in \emph{non-uniform} batch sizes among different workers, which may negatively affect the training accuracy. To ensure that model training process can still converge correctly even with inconsistent batch sizes, we further propose \emph{weighted gradient aggregation}, in which the batch size of each worker is used as the weight of its gradient in aggregation. 

We have implemented LB-BSP with a Python module that can be easily integrated into existing ML frameworks like TensorFlow~\cite{abadi2016tensorflow}, PyTorch~\cite{PyTorch} and MXNet~\cite{chen2015mxnet}. Our experiments on Amazon EC2 with popular benchmark training workloads show that LB-BSP can effectively eliminate stragglers in non-dedicated clusters, achieving near-optimal training efficiency with negligible overhead. In particular, in a 16-node heterogeneous GPU cluster, LB-BSP outperforms existing worker coordination schemes by over $54\%$; and in a 32-node shared CPU cluster, it attains an improvement of up to 38.7\% over state-of-the-art straggler mitigating approaches.

\section{Background and Motivation}
\label{sec:motivation}

\subsection{Research Background}
\label{sec:background}

\noindent\textbf{Basics of Machine Learning.} Given a machine learning (ML) model, the objective of model training process is to find the \emph{model parameters} $\omega^\star$ that can minimize the \emph{loss function} $L(\omega)$ over the labeled training dataset $S$, i.e.,
\begin{equation}
  \omega^\star=\arg\min_{\omega} {L(\omega)} = \arg\min_{\omega} 
  \frac{1}{|S|}\sum\nolimits_{s\in S}{l(s, \omega)}.
\end{equation}
Here, $l(s, \omega)$ is the loss value for a data sample $s$ in $S$. 

A popular training algorithm is \emph{mini-batch Stochastic Gradient Descent}~\cite{Dean12a,li2014efficient}, or simply SGD\footnote{For simplicity, we focus on standard SGD in this paper. Yet our LB-BSP solution can also be applied to all SGD variants (e.g., Adam or RMSProp, which make use of model gradients in different manners), because LB-BSP itself does not compromise the aggregated gradients (\cref{sec:preserving}).}. Its basic idea is to iteratively refine model parameters $\omega$ with the gradients $g$ calculated from the sample \emph{batches}, i.e.,
\begin{equation}
\omega^{k+1}=\omega^k-\eta g^k\textnormal{, where } g^k=\frac{1}{|B^k|}\sum\nolimits_{s\in B^k}{\nabla l(s, \omega^k)}.
\end{equation}
Here, $k$ is the iteration number, $B^k$ is a batch randomly sampled from the training dataset $S$, and $\eta$ is the \emph{learning rate}. 
Such training iterations would repeat until the model parameters $\omega$ finally converge.

Given the stochastic nature of SGD, the relationship between training completion (i.e., model convergence) and the sample processing amount is not deterministic, and the model training efficiency is usually decoupled into two parts: \emph{hardware efficiency}---how fast each iteration can be finished, and \emph{statistical efficiency}---how many iterations it takes towards model convergence. Therefore, efficient model training requires both high hardware efficiency and high statistical efficiency.

\vspace{.6em}
\noindent\textbf{Distributed Model Training.} To accelerate the model training process, it has been increasingly common to train ML models in a distributed manner, with a pool of parallel workers. Those workers typically collaborate in a synchronous mode (i.e., BSP)  \cite{zhang2017poseidon,goyal2017accurate,cui2016geeps}, under the \emph{Parameter Server} (PS)~\cite{ho2013more,Li14a,cui2016geeps} or \emph{All-Reduce}~\cite{goyal2017accurate} architecture. 

With the widespread adoption of ML techniques, distributed model training is now conducted in various cluster environments, and we broadly classify them into two types: \emph{dedicated} and \emph{non-dedicated} clusters.

\textbf{\emph{(1) Dedicated clusters}} implies that the clusters are homogeneous, and are composed of cutting-edge GPUs dedicated to one training job. For example, Facebook's well-known work of training ImageNet in one hour~\cite{goyal2017accurate} is conducted in such a dedicated cluster with 256 Tesla P100 GPUs. While dedicated clusters can yield high training efficiency, they are expensive to maintain and not widely accessible~\cite{goyal2017accurate,jia2018highly}.

\textbf{\emph{(2) Non-dedicated clusters}} refer to the clusters that are composed of heterogeneous hardware or shared by multiple tenants. Since newer CPU or GPU generations get released at a rapid pace, a budget-limited user may choose to train models with a set of heterogeneous hardware (e.g., GPUs) from the local inventory~\cite{li2018ease,park2020hetpipe} or from spot markets such as Amazon EC2~\cite{ec2_spot,gpu_spot,jiang2017heterogeneity} or FloydHub~\cite{floydhub}. Meanwhile, large companies~\cite{hazelwood2018applied,jeon2018multi} may host multiple ML jobs in their production clusters with mixed hardware types; for fairness, users may be allocated heterogeneous hardware~\cite{Chaudhary2020BalancingEA,Mahajan2020ThemisFA,narayanan2020heterogeneity}, and their resource allocation may also be dynamically adjusted by cluster schedulers for resource packing purpose~\cite{xiao2018gandiva,gu2019tiresias}. 

\vspace{.2em}
While model training in non-dedicated clusters is increasingly common, it is far less efficient than training in dedicated clusters due to the straggler problem, which is more severe in non-dedicated clusters. For clarity, we classify stragglers into two groups: \emph{non-deterministic} and \emph{deterministic} stragglers. 

\textbf{\emph{(1) Non-deterministic stragglers}} are caused by temporary disturbance like OS jitter or garbage collection, usually being transient and slight. They occur and vanish naturally in both dedicated and non-dedicated clusters. 

\textbf{\emph{(2) Deterministic stragglers}} are caused by the heterogeneity of worker resource quality or quantity, and are often severe and long-lasting. They occur only in non-dedicated clusters.

Compared with non-deterministic stragglers, deterministic stragglers are more harmful to distributed model training, by forcing fast workers to always wait for the slowest one in each iteration if under the popular BSP scheme. With the increasing prevalence of non-dedicated clusters, it is thus in urgent need to tame such deterministic stragglers.

\subsection{Related Work}
\label{sec:sync_schemes}

Stragglers have long been a thorny problem in distributed computing systems, and in the research literature there have been many attempts to tackle such a problem.

\subsubsection{Bypassing Stragglers with Relaxed Synchronization}\

\vspace{.2em}
\noindent To exempt fast workers from waiting for stragglers, two worker coordinating schemes with the synchronization constraint compromised have been proposed: ASynchronous Parallel (ASP), and Stale Synchronous Parallel (SSP).

\vspace{.4em}
\noindent\textbf{ASP.} In ASP~\cite{Dean12a}, workers can independently proceed to the next iteration without waiting for others. In this way, ASP wastes \emph{no} compute cycles and attains high hardware efficiency. However, the price paid is low statistical efficiency: without global synchronization, the gradient computation often uses \emph{stale} parameters, which yields low-quality updates and requires more iterations to converge~\cite{ho2013more,Langford09a}.

\vspace{.4em}
\noindent\textbf{SSP.} SSP~\cite{ho2013more,Cui14a} comes as a middle ground between BSP and ASP. In SSP, fast workers wait for the stragglers \emph{only when} the parameter \emph{staleness} reaches a particular threshold. SSP can then accelerate ML iterations while providing a convergence guarantee. Nonetheless, SSP focuses primarily on non-deterministic stragglers, expecting the straggling workers to catch up soon in later iterations. This works for homogeneous clusters, but  in non-dedicated clusters the deterministic stragglers may persist across many consecutive iterations, and the staleness quota of SSP can be quickly used up. After the quota is used up, fast workers will have to wait for the slowest ones in almost every iteration, leading to \emph{low hardware efficiency}.

\subsubsection{Mitigating Stragglers by Redundant Execution}
\label{sec:r-e}\

\vspace{.2em}
\noindent\textbf{\emph{Redundant execution}} \cite{ananthanarayanan2013effective,zaharia2008improving} is widely adopted to mitigate stragglers in traditional data analytics frameworks like MapReduce~\cite{dean2010mapreduce} and Spark~\cite{zaharia2012resilient}. It launches multiple copies of a straggling task and accepts results only from the one that finishes first. It has also been recently introduced to the context of distributed machine learning: Chen \emph{et al.}~\cite{chen2016revisiting} proposed to train deep neural models with extra backup workers and to use gradients from those workers finishing the earliest. However, redundant execution is merely a suboptimal solution: it mitigates some worst-case stragglers but fails to eliminate all of the stragglers \emph{completely}. Even worse, backup workers themselves would consume additional resources.

\subsubsection{Eliminating Stragglers by Load Balancing}
\label{sec:load-balancing}\

\vspace{.2em}
\noindent The solutions we have discussed so far are agnostic to the root causes of stragglers, and they do not seek to prevent stragglers from occurring. As we mentioned, efficiency loss in non-dedicated clusters is primarily caused by deterministic stragglers, whose root cause is very clear---the mismatch between the workers' loads and their processing capabilities (existing ML frameworks~\cite{abadi2016tensorflow,chen2015mxnet,jia2014caffe} blindly assign a \emph{uniform} batch load to all the workers). Therefore, to fundamentally eliminate such deterministic stragglers related to load-resource mismatching, we should resort to \emph{load balancing} techniques. Load balancing is a classical research topic~\cite{al2012survey} 
in the parallel processing community, and existing solutions can be broadly classified into two types: \emph{static} and \emph{dynamic} load balancing.

\vspace{.4em}
\noindent\textbf{Static Load Balancing.} Static load balancing strategies~\cite{kokilavani2011load,samal2013analysis,tang2000optimizing,tantawi1985optimal} set a constant load for each worker---as the execution commences---under a given scheme like Round-Robin~\cite{samal2013analysis} or with some static knowledge of the worker status~\cite{tang2000optimizing,tantawi1985optimal}. 
It does not require real-time progress measurements or cross-worker communication, but cannot react to resource variations that are common in non-dedicated clusters.

\vspace{.4em}
\noindent\textbf{Dynamic Load Balancing.}
Dynamic load balancing strategies use \emph{work-stealing} or \emph{work-shedding} to redistribute load from heavily-loaded workers to lightly-loaded ones at runtime~\cite{blumofe1999scheduling,dinan2009scalable,acar2013scheduling,yang2018scheduling,acun2016mitigating}. They are mostly developed for traditional \emph{task}-based parallel programming models in multi-core or HPC systems. Recently, FlexRR~\cite{harlap2016addressing} adopted such a dynamic strategy to tackle (non-deterministic) stragglers for ML workloads. It measures the workers' instantaneous progress at a fine granularity (100 checks per iteration); once a straggling worker lags behind others over a given threshold, it would yield certain sample processing load to a faster worker. Responding to dynamic changes of resources, dynamic load balancing schemes usually outperform static ones, but the cost paid is much higher computation and communication overhead (for progress monitoring, status collection and workload migration), which is not desirable in resource-intensive ML training. To reduce such overhead, FlexRR conducts straggler detection and load migration within designated \emph{worker groups}, which nonetheless leads to suboptimal load balancing performance due to its lack of global coordination.

Moreover, a key assumption of dynamic load balancing strategies is that, workloads shall be processed in a \emph{sequential} manner so that they can be arbitrarily split and transferred at runtime. However, this is not true for modern ML workloads. Training ML models is compute-intensive, and parallel-processing accelerators like GPUs are commonly used in practice, for which sequential processing is highly inefficient (\cref{sec:profile_gpu}). To fully exploit the power of such accelerators, mainstream ML frameworks wrap all the samples in a batch as a tensor matrix (e.g., a \texttt{Tensor} in TensorFlow/PyTorch, or an \texttt{NDArray} in MXNet), which is concurrently processed \emph{in a single round}. Given such \emph{all-or-nothing} processing, it is hard to measure fine-grained worker progress or adjust its load in the midst of an iteration, making dynamic load balancing simply infeasible.

\subsection{Semi-Dynamic Load Balancing}
\label{sec:semi-dynamic}

\noindent\textbf{Objectives.} Based on our discussions so far, our objective is to develop a load balancing strategy for ML workloads with the following properties:

\textbf{\textit{(1) Practicality.}} It should be compatible with the style of tensor-based processing in existing ML frameworks.

\textbf{\textit{(2) Effectiveness.}} It should be aware of the instantaneous worker execution status, and adjust the workers' load in a coordinated manner to attain the best load balancing effect.

\textbf{\textit{(3) Efficiency.}} It should not interfere with regular processing within each iteration, and should try to be light-weight by avoiding cross-worker data movement.

\vspace{.4em}
\noindent\textbf{Design Philosophy.} To meet these objectives, we propose a new load balancing strategy called \emph{semi-dynamic load balancing}. Tailored for ML workloads, its basic idea is to have workers' load be \emph{static} within each iteration but \emph{dynamic} across different iterations. In particular, we offload all load balancing operations---status measurements, straggler detection and load adjustment---at the \emph{iteration boundaries of BSP}, using the \emph{batch size} as a tool for load tuning. This strategy is feasible and can satisfy all of our design objectives, which we rationalize from the following three aspects:

\textbf{\textit{(1) Measuring worker status at iteration boundaries.}} Training iterations in modern ML frameworks are relatively short (in \emph{seconds} or even \emph{sub-seconds}, as we show later in Fig.~\ref{fig:motivation_relationship} and Fig.~\ref{fig:relationship_gpu}), and these iterations share a high similarity because an identical computation graph is used in each iteration. Therefore, the execution status in recent iterations is a valuable reference for that of the near future, relieving the need for costly intra-iteration progress measurements.

\textbf{\textit{(2) Detecting stragglers at BSP iteration boundaries.}} Under BSP there is a synchronization barrier at the end of each iteration, offering a natural opportunity to centralize all the workers' status information and optimize, with a global view, their load for the best load balancing effect.

\textbf{\textit{(3) Adjusting load by tuning the batch size at iteration boundaries.}} The batch size is a hyper-parameter that, as each iteration commences, dictates how many samples should be encapsulated into a tensor batch for processing in that iteration. It accurately controls a worker's load because each sample consumes identical compute cycles. Besides, given the stochastic (i.e., \emph{sample-insensitive}) nature of SGD, load migration can be realized by increasing the batch size of one worker and reducing that on another. This can avoid the communication overhead without compromising the training convergence.

In the next section, we show how the philosophy of semi-dynamic load balancing can be implemented in practice.
\section{LB-BSP}
\label{sec:solution}
In this section, we present \emph{Load-Balanced Bulk Synchronous Parallel} (LB-BSP), an integrated scheme atop BSP, for efficient distributed learning in non-dedicated clusters. 
We start with the problem formulation of LB-BSP, and then elaborate our solutions for CPU and GPU clusters, respectively.

\subsection{Problem Formulation}
\label{sec:problem}

In each model training iteration, given the sample batch, a worker first calculates a local gradient and then remotely refines the global model. Here we refer to the entire duration as the \emph{batch processing time}, denoted by $t$.
It can be divided into two parts: \emph{computation time ($t^p$)}, which measures the time taken to compute the gradient, and \emph{communication time\footnote{
		Communication and computation are partially overlapped~\cite{zhang2017poseidon,goyal2017accurate,cui2016geeps} in existing ML frameworks (e.g., TensorFlow, MXNet); for clarity, communication time in our definition excludes the overlapped periods.
}
($t^m$)}, which measures the time taken for data transmission.

In a nutshell, LB-BSP seeks to equalize $t$ on different workers by rightsizing their sample batches at the synchronization barriers.
This can be formulated as an optimization problem. Given $n$ workers with the initial batch size \.{x}, we want to find the worker batch sizes $\vec{x}=(x_1, x_2, ..., x_n)$ that can minimize the longest batch processing time among all workers, i.e.,
\begin{equation}
\label{eq:optimization}
\begin{aligned}
& \underset{\vec{x}=(x_1, x_2, ..., x_n)}{\text{min}} & & \underset{i\in\{1,2,...n\}}{\text{max}} t_i, \\
& \text{\qquad s.t.} & & t_i = t^p_i + t^m_i, \; i=1,\ldots,n; \\
& & &  t^p_i = \varGamma_i(x_i), \; i=1,\ldots,n; \\
& & & \sum\nolimits_{i=1}^{n}{x_i} = X = n\text{\.{x}}.\\
\end{aligned}
\end{equation}
Here $\varGamma_i(\cdot)$ is the function between worker-$i$'s computation time $t^p_i$ and its batch size $x_i$. The last constraint ensures that the total number of samples processed in each iteration remains the same as that in BSP. Table~\ref{tbl:notations} summarizes the important notations used in the paper.

\newcommand{\specialcell}[2][c]{%
	\begin{tabular}[#1]{@{}c@{}}#2\end{tabular}}
\begin{table}[t]
	\centering
	\renewcommand{\arraystretch}{1.1}
	\footnotesize
	\begin{tabular}{||c|l||c|l||}
		\hline
		$\!t$ & batch processing time & $x_i$ & batch size on worker-$i$\\
		$\!t^p$ & computation time & \.{x}  &  initial batch size \\
		$\!t^m$  & communication time & $X$& $n$\.{x} ($n$ is worker number)  \\
		$\!\varGamma(\cdot)$ & function between $t^p$ and $x$ & $v$  & sample processing speed\\
		\hline
	\end{tabular}
	\caption{Summary of important notations.}
	\label{tbl:notations}
	\vspace{-0.3in}
\end{table}

\vspace{.4em}
\noindent\textbf{Solution Overview.}  
Solving problem~\eqref{eq:optimization} poses different challenges to CPU and GPU clusters. In shared CPU clusters, $t_i$ increases linearly with $x_i$, but that ratio varies with the temporal resources. In GPU clusters, the relationship between $t_i$ and $x_i$ is non-linear and hard to profile at runtime. Based on their respective characteristics, we design an \emph{analytical} method that directly configures the optimal $\vec{x}$ for CPU clusters (\cref{sec:cpu}), and a \emph{numerical} method that iteratively approaches the optimal $\vec{x}$ for GPU clusters (\cref{sec:gpu}).

\subsection{LB-BSP in CPU Clusters}
\label{sec:cpu}
We summarize some typical scenarios where models are trained in CPU clusters without
accelerators like GPUs:

\textit{\textbf{(1) Non-neural-network Model Training.}} Many traditional ML applications like Support Vector Machines (SVM) or Logistic Regression (LR) demand less computing resources than neural networks. They are usually trained in CPU clusters~\cite{jiang2017heterogeneity,hazelwood2018applied,harlap2016addressing,zhang2017slaq}. 

\textit{\textbf{(2) Non-urgent Model Training.}} Non-urgent ML tasks may also be trained in CPU clusters opportunistically with leftover resources~\cite{Le2020AlloXCA,hazelwood2018applied}.
For example, to improve cluster utilization, Facebook trains some peripheral face recognition models with the off-peak portions of CPU servers in the diurnal cycle, where the CPU resources would otherwise be wasted~\cite{hazelwood2018applied}. 
 
We next present the techniques to fully exploit available resources in such non-dedicated CPU clusters to attain the best model training efficiency.

\subsubsection{Performance Characterization of CPU Workers}
\label{sec:profile_cpu}\

\begin{figure}[t]
	\vspace{-0.15in}
	\centering
	\subfloat
	{
		\includegraphics[width=0.222\textwidth]{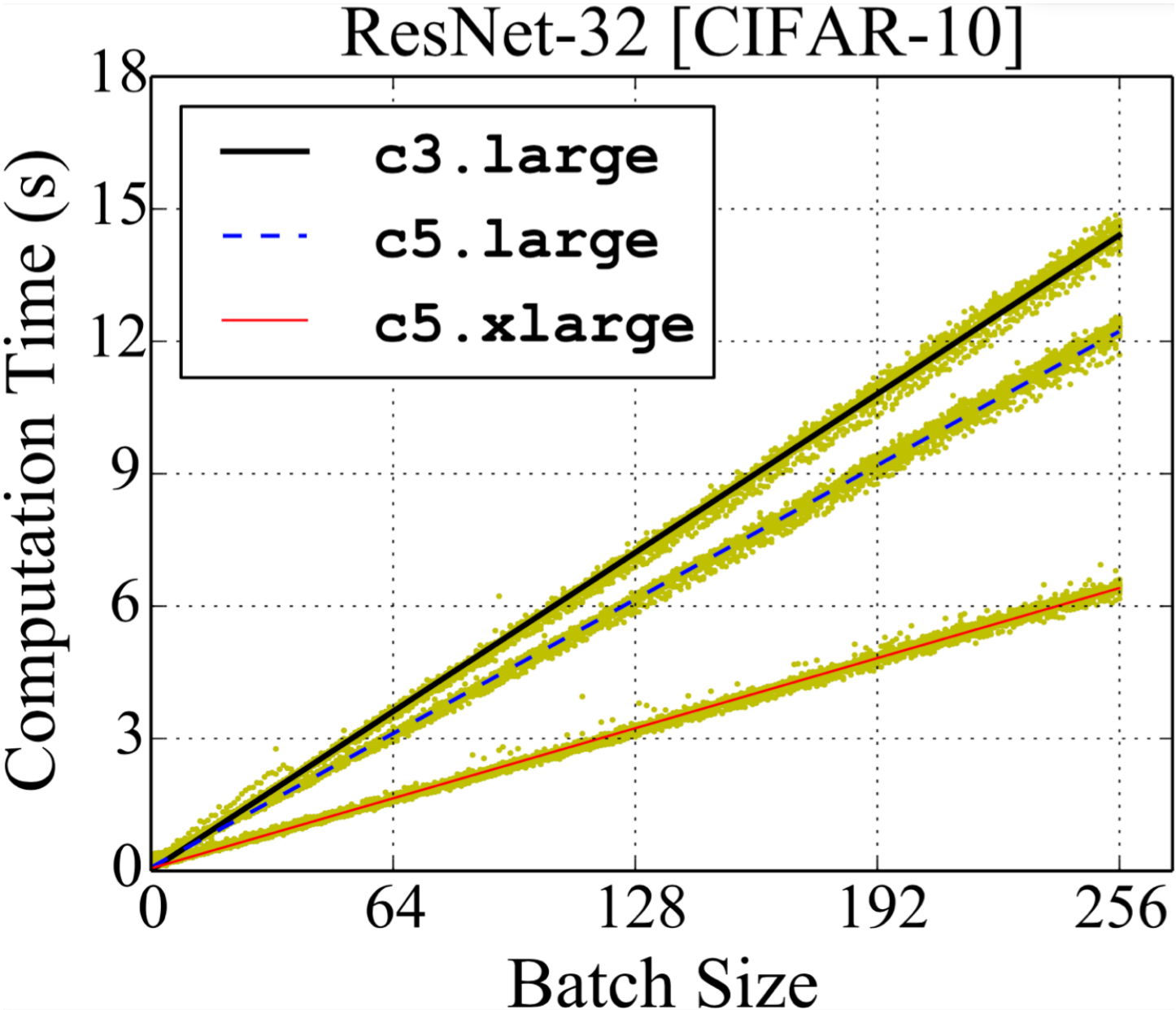}
	}\hfill
	\subfloat
	{	
		\includegraphics[width=0.225\textwidth]{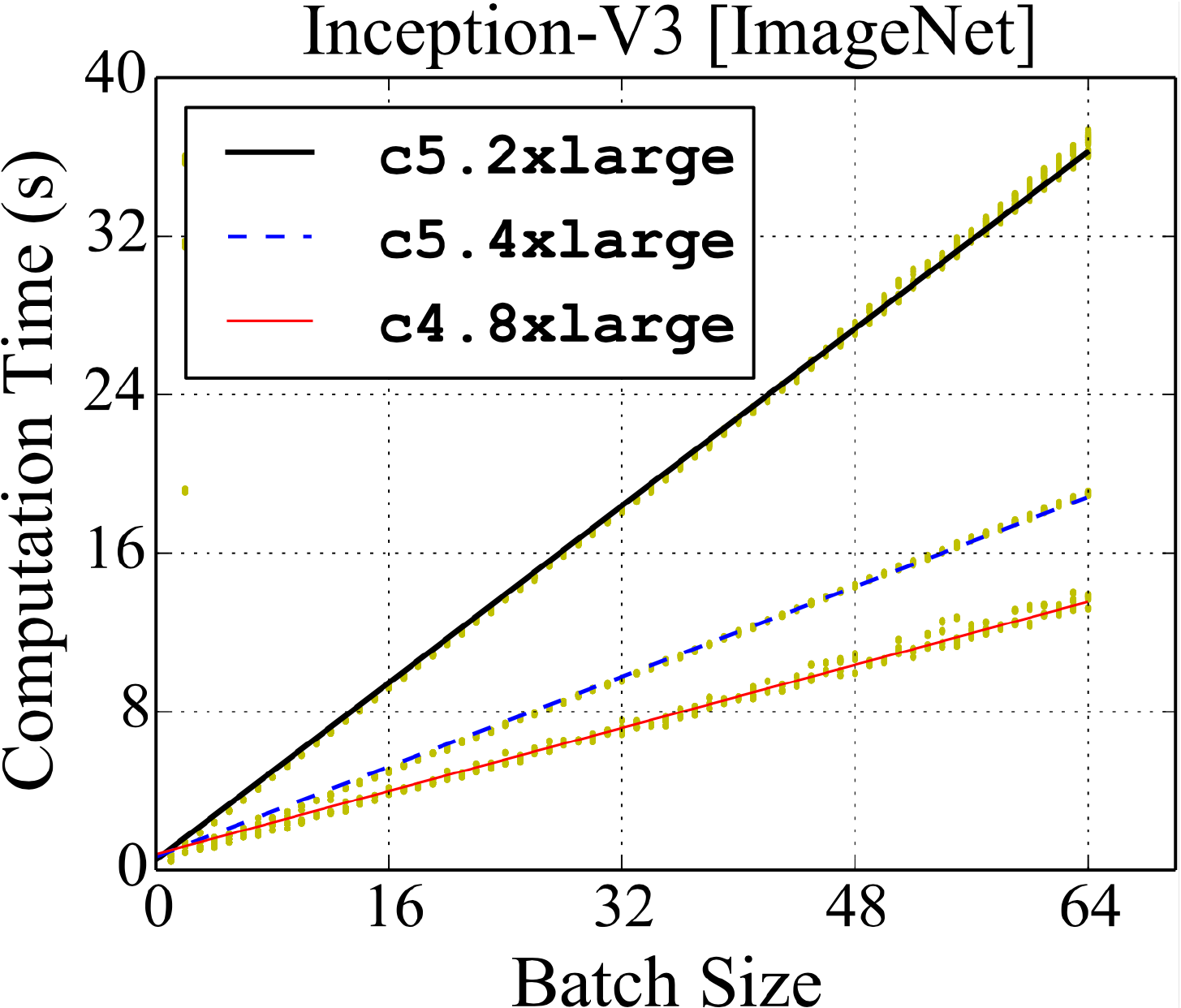}
	}
	\vspace{-0.1in}
	\caption{The relationship between computation time and batch size for different EC2 CPU instances.}
	\label{fig:motivation_relationship}
	\vspace{-0.2in}
\end{figure}

\vspace{.4em}
\noindent\textbf{Static Characteristics.} To solve problem~\eqref{eq:optimization} for CPU clusters, we first 
characterize the static properties of CPU workers by measuring their performance against different batch size configurations when there is no resource contention. 

\textit{\textbf{(1) Negligible Communication Time: $t^p\gg t^m$ ($t\approx t^p$).}} Model training is a compute-intensive task for CPU workers, and with built-in optimizations~\cite{zhang2017poseidon} of modern ML frameworks, communication can be hidden by the long computation time ($t^p$). 
In our measurements on EC2, when training the Inception-V3 model on ImageNet dataset (introduced later in \cref{sec:eval_gpu}) with 32 workers and one separate PS (\texttt{c5.2xlarge} instances with $\leq10Gbps$ bandwidth), $t^p$ takes more than $99\%$ of the batch processing time $t$. 
	
\textit{\textbf{(2) Linear Relationship: $\varGamma(x)=x/v.$}} As batch size quantifies the iteration load, for CPU processors the computation time $t^p$ is \emph{proportional} to the batch size $x$.
To confirm that, we respectively train the ResNet-32 and Inception-V3 model with different types of EC2 instances. We vary $x$ and record the corresponded computation time $t^p$ in Fig.~\ref{fig:motivation_relationship}, which in each case exhibits strong linearity between $x$ and $t^p$. Let $v$ be the ratio of $x$ to $t^p$, i.e., the \emph{sample processing speed}, we then have $t\approx t^p=\varGamma(x)=x/v$.

Given the characteristics above, we solve optimization problem~\eqref{eq:optimization} with $x_i=\frac{v_i}{\sum\nolimits_{j=1}^{n}{v_j}}X$. Note that such an analytical method essentially merges straggler detection and elimination together. In a nutshell, to set the worker batch size for an upcoming iteration, we only need to know their sample processing speed in that iteration. 

\vspace{.4em}
\noindent\textbf{Challenges for LB-BSP in shared CPU clusters.} Although a CPU worker's batch size can be determined with its sample processing speed, when training with resource contention in shared clusters, that sample processing speed may vary dynamically. Therefore, to load-balance workers with adjustments only at the iteration boundaries, we need to predict their sample processing speed before the start of an iteration. As stragglers in non-dedicated CPU clusters can be deterministic and non-deterministic, an ideal prediction approach should be robust to non-deterministic perturbations to avoid over-reaction or oscillation. It should also react to deterministic resource variations quickly.

\subsubsection{Predicting Sample Processing Speed}
\label{sec:narx}\

\begin{figure}[t]
	\centering
	\subfloat
	{
		\includegraphics[width=0.222\textwidth]{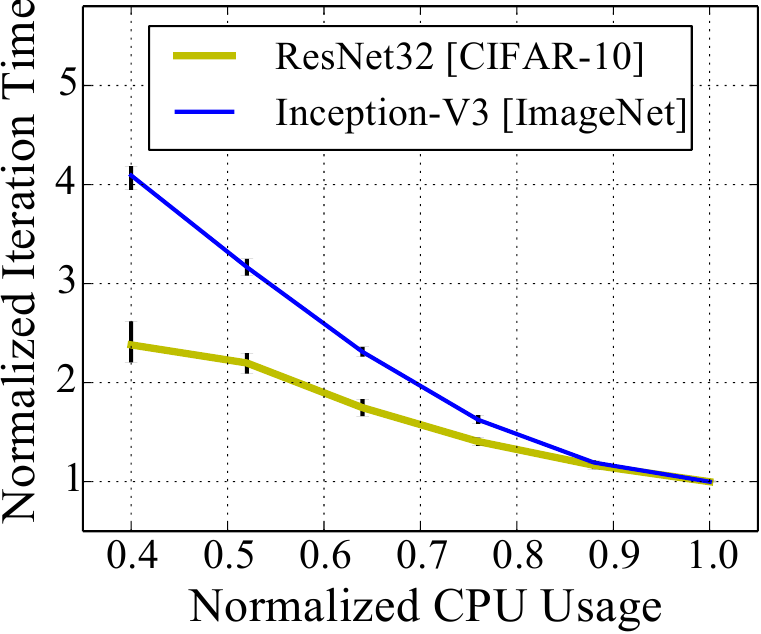}
	}\hfill
	\subfloat
	{	
		\includegraphics[width=0.222\textwidth]{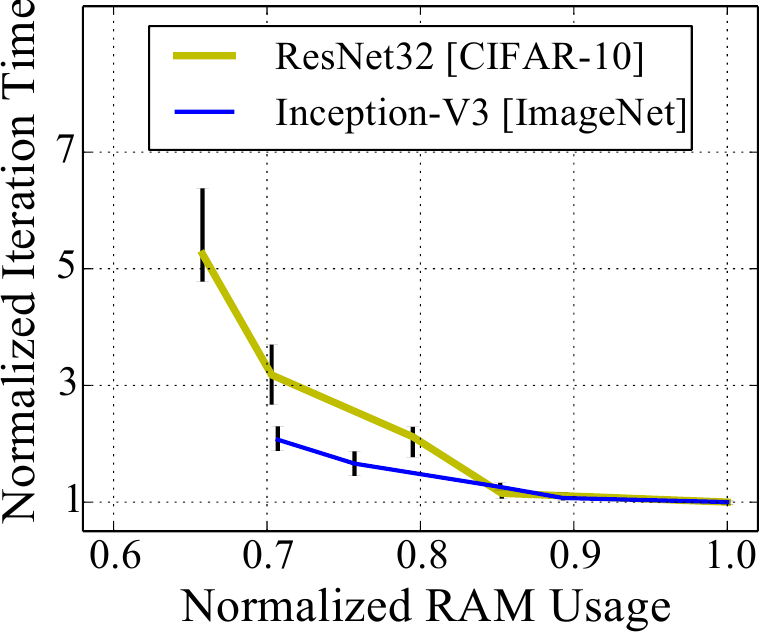}
	}
	\vspace{-0.1in}
	\caption{Sample processing speed is affected by CPU or memory resources. Measurements are conducted with the \texttt{stress-ng} tool~\cite{stress-ng} on a \texttt{c5.2xlarge} instance (with swap spaces enabled). The resource usage and iteration time are normalized by the monopolizing case. Each point is an average of $100$ iterations. 
	}
	\vspace{-0.15in}
	\label{fig:slowdown_factors}
\end{figure}

\vspace{.4em}
\noindent\textbf{Potential Approaches.}
A simple solution is to use the last iteration's speed or the Exponential Moving Average (EMA) speed as the predicted one. 
However, the former is not robust to temporary perturbations, and the latter cannot react quickly to drastic resource variations.

Speed prediction belongs to a classical research problem---\emph{time series prediction}, for which many \emph{statistical} or \emph{learning-based} techniques have been proposed. 
As a statistical approach, \emph{Autoregressive Integrated Moving Average} (ARIMA) \cite{ediger2007arima} makes predictions with statistics like average and deviation. Meanwhile, models based on \emph{Recurrent Neural Network} (RNN)~\cite{connor1994recurrent} (like plain RNN and LSTM~\cite{hochreiter1997long}), with the ability to maintain inner memory, have been applied in forecasting real-world time series like stock price~\cite{rather2015recurrent} or transport flow~\cite{polson2017deep}. 

However, the prediction performance of all the above approaches is limited by their \emph{blindness} to the underlying resources. 
In fact, variations of the driving resources like CPU and memory\footnote{
	Other resources (e.g., disk I/O, heat) may also impact sample processing speed. Yet instead of exhaustively exploring all the potential impact factors, our goal here is to show the benefits of including the driving resources in prediction, and CPU and memory are two easy-to-measure factors for that.
} closely relate to the instantaneous worker processing capability. 

This is supported by Fig.~\ref{fig:slowdown_factors}, in which the model training processes are slowed down after we restrict their resource usage. 
Such driving resources can help to distinguish the deterministic straggling factors from random perturbations and make more accurate prediction. To this end, we find that \emph{Nonlinear AutoRegressive eXogenous} (NARX) model \cite{gao2003narmax,diaconescu2008use} is a good fit for the prediction of sample processing speed.

\begin{figure}[t]
	\centering
	\subfloat
	{
		\includegraphics[width=0.38\textwidth]{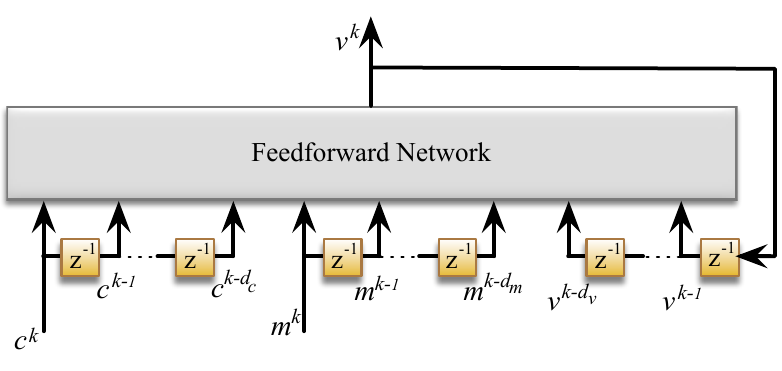}
	}
	\vspace{-0.15in}
	\caption{NARX architecture.}
	\label{fig:narx}
	\vspace{-0.15in}
\end{figure}

\vspace{.4em}
\noindent\textbf{NARX Approach.} The NARX model we use is basically an extended recurrent neural network that takes three series as inputs: past values of sample processing speed ($\boldsymbol{v}$), current and past values of the two \emph{driving} resources---CPU and memory usage ($\boldsymbol{c}/\boldsymbol{m}$). Essentially, NARX aims to learn a \emph{nonlinear function} $F(\cdot)$ between the predicted speed and a limited view (specified with a \emph{look-back window size}) of the input series: 
\begin{equation}
\!\!v^k=F(v^{k-1},...,v^{k-d_v},c^k,...,c^{k-d_c},m^{k},...,m^{k-d_m}).\!\!
\end{equation}
Here $v^k$, $c^k$ and $m^k$ represent the value of \emph{speed}, \emph{CPU} and \emph{memory} usage in iteration $k$, respectively; $d_v$, $d_c$ and $d_m$ represent the corresponded \emph{look-back window size} of each series. Fig.~\ref{fig:narx} shows the unfolded architecture of the NARX model, in which the input series are fed into a \emph{feedforward} neural network.

The batch size adjusting process with NARX is elaborated in Alg.\ref{alg:cpu-lbbsp}. 
In practice, to ensure high prediction accuracy, we maintain a NARX model for each worker, which is trained with the historical execution information (i.e., $\boldsymbol{v}$, $\boldsymbol{c}$ and $\boldsymbol{m}$). 
To avoid high model complexity, as in existing prediction works \cite{argyropoulos2016narx, cadenas2016wind}, the \emph{look-back window} sizes for all the three input series are set to $2$, and we include only \emph{one} hidden layer in the feedforward network. 
Such a simple model
can avoid over-fitting and converge fast. Our later evaluation (\cref{sec:deepdive}) confirms that such a NARX-based approach can make better predictions than other approaches we surveyed. 

\begin{algorithm}[t]
	\small
	\caption{Batch Size Updating in CPU clusters}
	\label{alg:cpu-lbbsp}
	\begin{algorithmic}[1]
		\Statex \hspace{-0.2in}\textbf{Input}: \{$x_i^{k-1}\!$\}, \{$t_i^{k-1}\!$\}, \{$c_i^{k}\!$\}, \{$m_i^{k}\!$\} \Comment $\!$\footnotesize last batch size, last batch processing time, current cpu \& memory usage of all the workers $\!(i\!\!=\!\!1,\!2,\!\!...,\!n)\!$ \small
		\Statex \hspace{-0.2in}\textbf{Require}: 
		past values of $\{v_i\}, \{c_i\}, \{m_i\}, i=1,2,...,n$;
		\Statex \hspace{0.28in} 
		$F_i(\cdot)$, $i=1,2,...,n.$ \Comment \footnotesize NARX model for each worker \small
		\Procedure{CPU\_UpdateBatchSize}{$k$}
		\State $\!\!\!v_i^{k-1}\leftarrow x_i^{k-1}/t_i^{k-1}, i$=$1,2,...,n.$
		\vspace{0.01in}
		\State $\!\!\!v_i^k\!\!=\!F_i(v_i^{k-1}\!\!,...,v_i^{k-d_v}\!,c_i^{k}\!,...,\!c_i^{k-d_c}\!\!,m_i^{k}\!,...,m_i^{k-d_m})$, $i$=$1,...,n.$
		\State $\!\!X\leftarrow \sum\nolimits_{i=1}^{n}{x_i^{k-1}}$
		\vspace{0.01in}
		\State $\!\!x_i^k=\frac{v_i^k}{\sum\nolimits_{j=1}^{n}{v_j^k}}\cdot X$, $i=1,2,...,n.$
		\State round $x_i^k (i=1,...,n)$ to integers with $\sum\nolimits_{i=1}^{n}{x_i^k}=X$.
		\State \Return \{$x_i^k$\}
		\EndProcedure
	\end{algorithmic}
	\normalsize
\end{algorithm}

\subsection{LB-BSP in GPU Clusters}
\label{sec:gpu}

With strong parallel processing capability, GPUs are the workhorse hardware for training deep neural networks~\cite{zhang2017poseidon,goyal2017accurate,jeon2018multi}. 
As elaborated in \cref{sec:background}, neural network models may be trained in non-dedicated GPU clusters: with heterogeneous GPU instances from the spot markets~\cite{floydhub,gpu_spot}, or in shared GPU clusters~\cite{jeon2018multi,peng2018optimus,xiao2018gandiva,gu2019tiresias} where workers may inter-connect at different locality levels and be migrated across machines for resource consolidation.
Nonetheless, implementing LB-BSP in non-dedicated GPU clusters renders a challenge different with that in CPU clusters. In this part, we first characterize the performance of standalone GPU workers, and then present our LB-BSP algorithm for GPU clusters.

\subsubsection{Performance Characterization of GPU Workers}
\label{sec:profile_gpu}\

\vspace{.4em}
\noindent\textbf{Static Characteristics.} 
To solve problem~\eqref{eq:optimization} for GPU clusters, we first profile the static properties of GPU workers. 

\textit{\textbf{(1) Non-negligible Communication Time: $t^p\not \gg t^m$ ($t\not \approx t^p$).}} Computations on GPU workers are usually orders of magnitude faster than CPU workers. Therefore, the communication time in each iteration is no longer negligible when compared with the computation time~\cite{zhang2017poseidon}. 

\begin{figure}[t]
	\centering
	\subfloat
	{
		\includegraphics[width=0.222\textwidth]{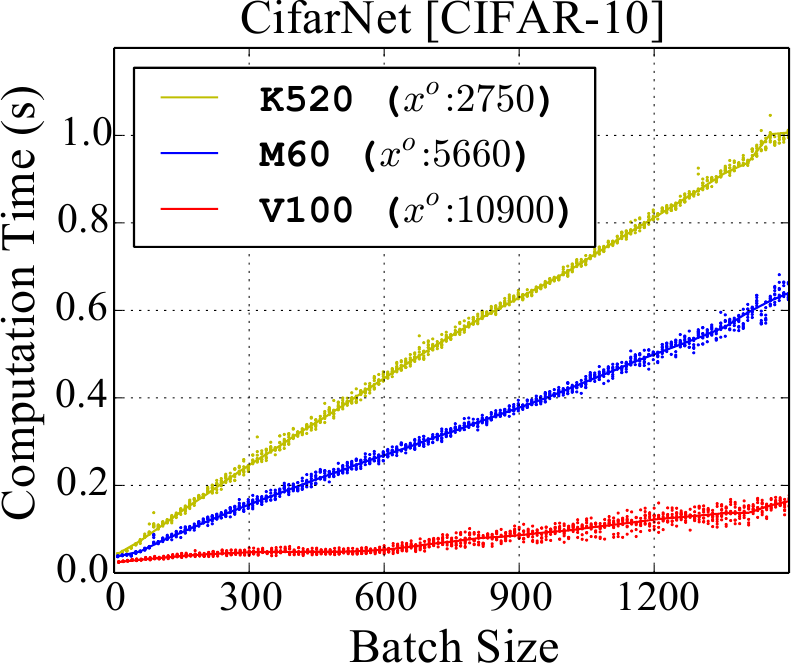}
	}\hfill
	\subfloat
	{	
		\includegraphics[width=0.213\textwidth]{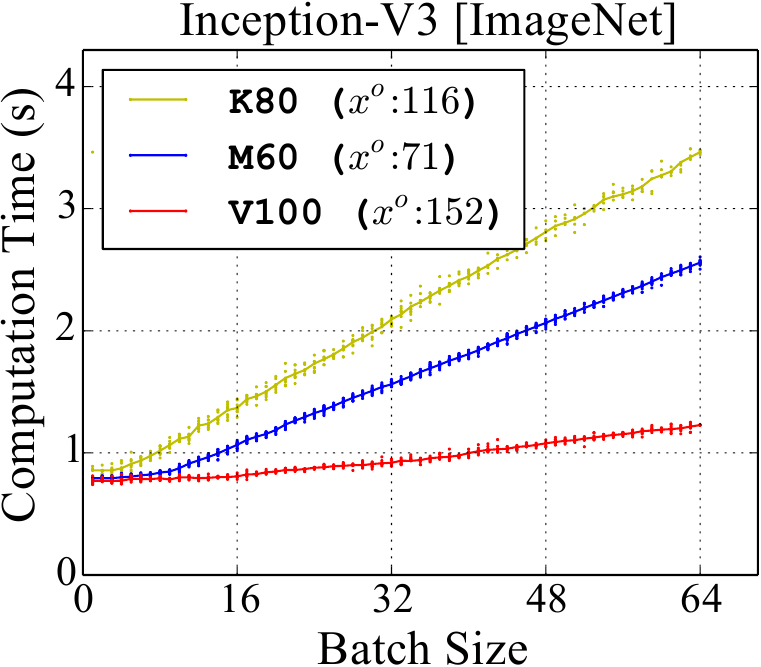}
	}
	\vspace{-0.1in}
	\caption{The relationship between computation time and batch size of different GPU types.} 
	\label{fig:relationship_gpu}
		\vspace{-0.1in}
\end{figure}

\textit{\textbf{(2) Non-negligible GPU Launching Overhead: $\varGamma(0)>0$.}} Fig.~\ref{fig:relationship_gpu} shows the relationship\footnote{
	To eliminate network interference, the curve for each GPU type is obtained with a TensorFlow worker process and a collocated PS process. The models we run are CifarNet (a CNN-based neural network~\cite{krizhevsky2009learning} designed to classify CIFAR-10 dataset) and Inception-V3.}
 $\varGamma(\cdot)$ between $t^p$ and $x$ for different GPU types. 
 Even for a very small batch, we find that the GPU computation time could still be considerable. This is because a GPU needs to do a series of preparation work~\cite{lustig2013reducing} to process each batch (e.g., exchanging parameters between GPU and CPU memory, launching processing kernels), which incurs considerable time overhead regardless of the batch size, and that overhead is particularly salient for large models like Inception-V3. 
 In particular, the existence of GPU launching overhead confirms that \emph{sequential sample processing is highly inefficient}: it is unacceptable to sustain such an overhead when processing each sample.

\textit{\textbf{(3) GPU Saturation Effect: $\varGamma(x)=C, \forall x<x^s$.}} For advanced GPUs like Tesla V100, the batch computation time $t^p$ would be almost a \emph{constant} if the sample batch is too small (less than the saturation threshold $x^s$) to \emph{saturate} all the processing kernels~\cite{magni2014exploiting,li2016gpu}. Reducing batch size under $x^s$ does not help to reduce the computation time and would cause GPU underutilization. 

\textit{\textbf{(4) GPU Memory Limitation: $x<x^o$.}} During each training iteration, the input \emph{sample batch}, \emph{model parameters} and \emph{intermediate results} shall all reside in GPU memory. Therefore, the GPU memory size imposes a limit $x^o$ on the maximum batch size (as marked in the legends of Fig.~\ref{fig:relationship_gpu}), which, to avoid OOM (out-of-memory) error, must be complied with when adjusting the batch size of a GPU worker.

\vspace{.4em}
\noindent\textbf{Challenges for LB-BSP in GPU clusters.} The challenges for realizing LB-BSP in GPU clusters are quite different from that in CPU clusters. 
First, fine-grained GPU sharing is quite rare (due to the technical difficulty and overhead~\cite{gu2017deepprof,jeon2018multi}), and auxiliary resources (e.g., CPU, memory, network connectivity) provisioned in GPU clusters fluctuate less often than in CPU clusters~\cite{jeon2018multi,peng2018optimus,xiao2018gandiva,gu2019tiresias}, leaving it unnecessary to make real-time performance predictions for GPU workers.
Second, however, statically profiling the non-linear relationship $\varGamma(\cdot)$ incurs non-trivial programming and time overhead, and is particularly inconvenient for shared GPU clusters where the workers of a ML job may be migrated from time to time. 
Third, Fig.~\ref{fig:relationship_gpu} implies that batch processing time increases \emph{monotonically} with the batch size; meanwhile, compared with the huge number of seconds-level short iterations in the long training process, the occurrence of job migration
is much fewer, suggesting that the worker performance is stable in most consecutive iterations. 

Therefore, instead of analytically solving problem~\eqref{eq:optimization} based on static profiling, for GPU clusters it is more appropriate to employ a \emph{numerical approximation} method.

\subsubsection{A Drop-in Algorithm for LB-BSP in GPU Clusters}\

\vspace{.4em}
\noindent In this part, we devise a drop-in algorithm to iteratively approximate the \emph{equilibrium} where the gap among all the workers' batch processing time is minimized. 

The whole batch size adjusting algorithm is elaborated in Alg.~\ref{alg:gpu-lbbsp}. After each training iteration, we identify two GPU workers: a \emph{leader}---the GPU worker with the \emph{shortest} batch processing time, and a \emph{straggler}---the GPU worker with the \emph{longest} batch processing time. If the \emph{leader} has consecutively preceded the \emph{straggler} during an \emph{observation window} (observation window is introduced for robustness to non-deterministic random variations), we respectively increase (reduce) the \emph{leader} (\emph{straggler})'s batch size by a certain amount called \emph{step size}. Note that any GPU worker with less than 5\% available memory would, to avoid OOM error, be excluded from being identified as the \emph{leader}. Moreover, if with Alg.~\ref{alg:gpu-lbbsp} a worker's batch size is to be reduced below $0$, we should restart the training process without that worker, and the resultant setup is actually more efficient. 

Moreover, to reduce resource wastage, the equilibrium should be approached efficiently with minimum oscillation. To this end, we introduce two phases: a \emph{fast-approach} phase and a \emph{fine-tune} phase. Initially the algorithm enters the fast-approach phase, where we set a relatively large step size and short observation window; then, once oscillation---a former \emph{leader} now identified as a \emph{straggler} or vice versa---is detected, we switch to the fine-tune phase by reducing the step size (e.g., to $1$) and increasing the observation window size. 

\vspace{.4em}
\noindent\textbf{Remarks.} While Alg.~\ref{alg:gpu-lbbsp} is designed for GPU clusters, it also applies to other accelerators like TPU or FPGA. Existing measurements have shown that TPU~\cite{kochura2018batch} and FPGA~\cite{posewsky2018throughput} also exhibit a \emph{non-linear}, \emph{monotonically-increasing} relationship between $t^p$ and $x$, making it feasible to adopt LB-BSP in heterogeneous clusters with those hardware. 
	
\begin{algorithm}[t]
	\small
	\caption{Batch Size Updating in GPU Clusters}
	\label{alg:gpu-lbbsp}
	\begin{algorithmic}[1]
		\Statex \hspace{-0.2in}\textbf{Input}: \{$x_i^{k-1}$\}, \{$t_i^{k-1}$\}, \{$m_i^{k}$\} \Comment \footnotesize last batch size, last batch processing time \& current GPU memory usage of all workers $(i\!=\!1,...,n)$ 
		\small
		\Statex \hspace{-0.2in}\textbf{Require}:
		past values of $\{t_i\}, i=1,2,...n$;
		\Statex \hspace{0.28in} 
		$\!\varDelta\!\leftarrow\!5$, $D\!\leftarrow\! 5$ \Comment \footnotesize $\varDelta$: step size; $D$: observation window size\small
		\Procedure{GPU\_UpdateBatchSize}{$k$}
		\State \textit{leader} $\leftarrow$ $\arg\min_i\{t_i^{k-1}\mid m_i^{k}\leq 0.95\}$
		\State \textit{straggler} $\leftarrow$ $\arg\max_i\{t_i^{k-1}\}$
		\State $x_i^k\leftarrow x_i^{k-1}, i=1,2,...,n$
		\If {$x_\text{\textit{straggler}}^{k-1}\leq \varDelta$}
		\State \Return \{$x_i^{k}$\} \Comment \footnotesize print warning: remove this \textit{straggler}!
		\EndIf
		\If {$\forall h \in \{k-D, ..., k-1\},t_\text{\textit{leader}}^h<t_\text{\textit{straggler}}^h$} 
		\State $x_\text{\textit{leader}}^{k}\leftarrow x_\text{\textit{leader}}^{k-1}+\varDelta$
		; $x_\text{\textit{straggler}}^{k}\leftarrow x_\text{\textit{straggler}}^{k-1}-\varDelta$
		\vspace{0.02in}
		\ElsIf {$\exists h\in \{1, ..., k-1\}, t_\text{\textit{leader}}^h>t_\text{\textit{straggler}}^h$}
		\vspace{0.02in}
		\State  $\varDelta\leftarrow 1$, $D\leftarrow 20$ \Comment \footnotesize switch to \emph{fine-tune} phase \small
		\EndIf 
		\Return \{$x_i^{k}$\}
		\EndProcedure
	\end{algorithmic}
	\normalsize
\end{algorithm}
\subsection{Weighted Gradient Aggregation}
\label{sec:preserving}
By tuning the worker batch size with Alg.~\ref{alg:cpu-lbbsp} and Alg.~\ref{alg:gpu-lbbsp}, we can effectively eliminate deterministic stragglers and achieve high hardware efficiency. 
Yet, the resultant batch sizes on different workers would be inconsistent, which may affect the statistical efficiency. In this part, we first elaborate that problem and then give our solution.

\vspace{.4em}
\noindent\textbf{Problem of Naive Aggregation under LB-BSP.}
Under BSP, the aggregated global gradient $g$ for parameter updating is the naive average of the gradients from all the workers. Suppose there are $n$ workers and $g_i$ is the gradient calculated on worker-$i$ ($i$=$1,2,...,n$), then
\begin{equation}
g=\frac{1}{n}\sum_{i=1}^{n}g_i, \textnormal{  where } g_i=\frac{1}{|B_i|}\sum_{s\in B_i}{\nabla l(s, \omega)}.
\end{equation}
Here $B_i$ is the batch on worker-$i$. Getting rid of $g_i$, we have
\begin{equation}
g=\frac{1}{n}\sum_{i=1}^{n}\frac{1}{|B_i|}\sum_{s\in B_i}{\nabla l(s, \omega)}=\sum_{i=1}^{n}\sum_{s\in B_i}{\frac{1}{n|B_i|}\!\!\cdot\!\nabla l(s, \omega)}.\!
\end{equation}
This implies that, when workers have different batch sizes ($|B_i|$), the \emph{significance} of different samples, i.e., $\frac{1}{n|B_i|}$, is also different. Thus $g$ is biased to samples in small batches, which may harm the statistical efficiency. 
To verify that, we train the Inception-V3 model under Alg.~\ref{alg:gpu-lbbsp} in a 16-node heterogeneous GPU cluster (i.e., Cluster-A in \cref{sec:eval_gpu}). After traversing the ImageNet dataset for 20 epochs, the training accuracy only reaches 43\%, much worse than that under BSP (59\%). 

\vspace{.4em}
\noindent\textbf{Weighted Gradient Aggregation.} To avoid biased gradient, we propose \emph{weighted gradient aggregation}---using a worker's batch size as the weight when aggregating gradients. Suppose the total batch size is $\sum\nolimits_{j=1}^n{|B_j|}\!=\!\!X$, then we have
\begin{equation}
g=\frac{1}{\sum\nolimits_{j=1}^n{|B_j|}}\sum_{i=1}^{n}{|B_i|\cdot g_i}=\sum_{i=1}^{n}\sum_{s\in B_i}{\frac{1}{X}\cdot\nabla l(s, \omega)}.
\end{equation}
Obviously, now each sample plays an equal role in parameters updating, regardless of the batch size inconsistency. After that fix, for \emph{i.i.d.} dataset, i.e., samples being \emph{independent} and \emph{identically distributed}, LB-BSP can achieve identical statistical efficiency with BSP, which is known to be optimal. 

\subsection{Discussion on Data  Access Frequency}
\label{sec:discussion}
In this paper, we focus on cases where each worker can access the \emph{entire} dataset via local storage or Network File System (NFS)~\cite{shvachko2010hadoop,glusterfs,schmuck2002gpfs}. In shared production clusters, it has become almost a norm to store data in NFS (the access delay can be made negligible via pre-fetching), which largely facilitates data management  and job migration~\cite{hazelwood2018applied,jeon2018multi,xiao2018gandiva,gu2019tiresias}.

Nonetheless, there may still exist some cases that each worker can only access a local partition of the training dataset. Under LB-BSP where faster workers iterate with larger batches, this means that samples on faster workers would be accessed more frequently. Such a problem of \emph{uneven sample access frequency} may lead to inaccurate training results, especially when the dataset is not well shuffled before being partitioned and there are huge gaps on worker processing capability.
To address uneven sample access frequency with transparency to the upper level training process, we suggest an iterative SSP-style data scheduling scheme: once the traversal times of one partition exceed another over a given \emph{threshold}, we launch a background process to migrate certain amount of samples from the slower worker to the faster one. We have prototyped\footnote{
	We develop a Python module called \texttt{DataPartitionManager} to periodically collect each partition's traversal status and launch peer-to-peer data transmission when appropriate, where all the communications are in Thrift RPC calls. Once a data shifting process finishes, we reset the input stream (e.g., \texttt{DataLoader} in PyTorch) and partition-traversal statistics on all the workers. In our verification, we train the ResNet-32 model with 5 workers; each worker hosts two classes of the CIFAR-10 dataset and their batch sizes are 64, 96, 128, 160 and 192, respectively. With the \texttt{DataPartitionManager} and the gap of traversal times bounded by 2, we obtain the same test accuracy (0.92) as training with \emph{i.i.d} dataset.}
this method atop PyTorch,
and verified that it could make ideal accuracy even under \emph{non-i.i.d.} data distribution.
Such a kind of data migration method is indeed light-weight in GPU clusters, because in GPU clusters the worker speed is relatively stable and data migration is done-once-and-working-forever, with the overhead amortized. 
\section{Implementation}
\label{sec:implementation}
\begin{algorithm}[t]
	\small
	\caption{LB-BSP Workflow}
	\label{alg:1}
	\begin{algorithmic}[1]
		\Statex \hspace{-0.2in}\underline{\textbf{Worker: i=1, 2, ..., n:}}
		\Procedure{WorkerIterate}{$k$}
		\State $x_i\!\leftarrow\!$ \texttt{BatchSizeManager}.UpdateBatchSize($<$states$>$) 
		\Statex \Comment \footnotesize \emph{blocking} in CPU clusters and \emph{non-blocking} in GPU clusters\small
		\State load the next data batch $B_i^k$ such that $|B_i^k|=x_i$
		\State pull $w^{k}$ from PS
		\State calculate local gradient $g_i^k$ 
		\State push $\bar{g}_i^k\leftarrow \frac{nx_i}{X}g_i^k$ to PS \Comment \footnotesize$\bar{g}_i^k$: weighted gradient \small
		\EndProcedure
	\end{algorithmic}
	\begin{algorithmic}[1]
		\Statex \hspace{-0.2in}\underline{\textbf{Parameter Server (PS):}}
		\Procedure{ParameterServerIterate}{$k$}
		\State aggregate gradient $g^k$ $\leftarrow$$\frac{1}{n}\sum_{i=1}^{n}\bar{g}_i^k=\frac{1}{X}\sum_{i=1}^{n}|B_i^k|g_i^k$
		\State update parameters $w^{k+1}$ $\leftarrow$ $w^k-\eta g^k$ \Comment \footnotesize $\eta$: learning rate \small
		\EndProcedure
	\end{algorithmic}
	\begin{algorithmic}[1]
		\Statex \hspace{-0.2in}\underline{\textbf{BatchSizeManager:}}
		\Procedure{UpdateBatchSize}{$<$states$>$}
		\State Refer to Alg.~\ref{alg:cpu-lbbsp} (CPU cluster) or Alg.~\ref{alg:gpu-lbbsp} (GPU cluster).
		\EndProcedure
	\end{algorithmic}
	\normalsize
\end{algorithm}
We implement LB-BSP with  \texttt{BatchSizeManager}\footnote{
	While LB-BSP can be integrated into the engines of ML frameworks, this would lose generality,
	mess the decoupled programming logic of input and graph propagation modules, and also lose the flexibility to switch to All-Reduce communication backend (Operations in Alg.~\ref{alg:cpu-lbbsp} and Alg.~\ref{alg:gpu-lbbsp} are hard to be realized with All-Reduce semantics). 
}, a Python module
that can be integrated into existing ML frameworks including TensorFlow~\cite{abadi2016tensorflow}, PyTorch~\cite{PyTorch} and MXNet~\cite{chen2015mxnet}.

\begin{figure}[t]
	\centering
	\subfloat
	{
		\includegraphics[width=0.33\textwidth]{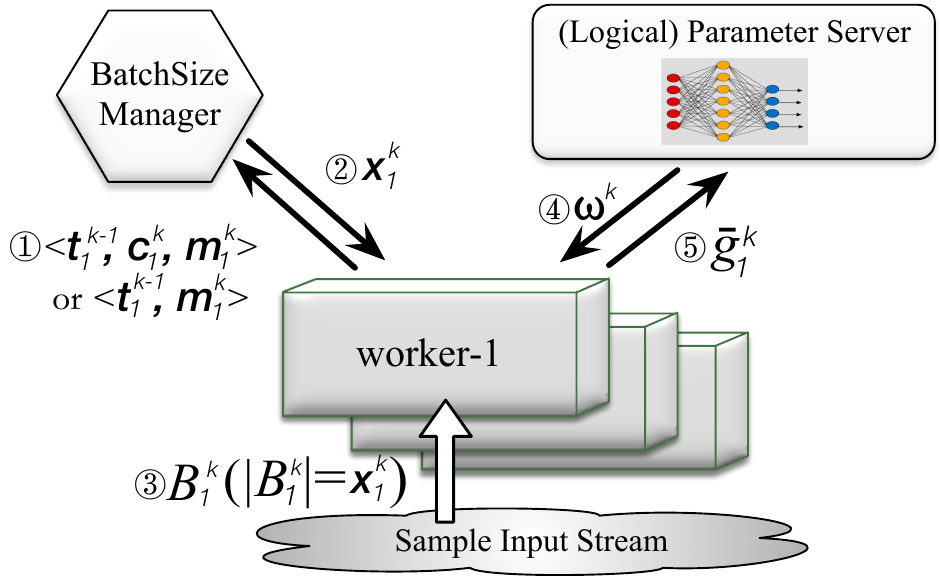}
	}
	\vspace{-0.1in}
	\caption{LB-BSP workflow in iteration $k$ (the circled numbers represent the execution order). The logical PS can be in the form of distributed shards, or be replaced by the \texttt{All-Reduce} architecture.}
	\vspace{-0.2in}
	\label{fig:framework}
\end{figure}

\vspace{.5em}
\noindent\textbf{Architecture Overview.} The overall workflow of LB-BSP is described in Alg.~\ref{alg:1} and Fig.~\ref{fig:framework}.
In the beginning of iteration $k$, each worker pushes its latest execution information 
($\langle$ batch processing time $t^{k-1}$, CPU usage $c^k$, memory usage $m^k$ $\rangle$ for CPU clusters, or $\langle t^{k-1}$, $m^{k}\rangle$ for GPU clusters)
to the \texttt{BatchSizeManager}, and then pulls back the updated batch size $x^k$. 
 Note that LB-BSP is also applicable if the gradients are aggregated with the All-Reduce architecture.

In particular, for CPU clusters the batch size updating process
is \emph{blocking} so that the \texttt{BatchSizeManager} can get the latest state information for speed prediction; yet for GPU clusters, it is \emph{non-blocking} because GPU workers' state is more stable (\cref{sec:gpu}), and non-blocking interaction can avoid prolonging the very short GPU iterations. 
To that end, on each GPU worker we launch a separate thread to update batch size in the background. 

\vspace{.4em}
\noindent\textbf{Enabling Variable Batch Size.} In existing ML frameworks, batch size is set as a constant when defining the computation graph, with no direct APIs to configure it during the training process. To enable variable batch size, in TensorFlow we decouple the batch size from the symbolic dataflow graph, and specify it as a \texttt{tensor} value passed to TensorFlow \texttt{session} through the \texttt{feed\_dict} API. For MXNet and PyTorch, we respectively customize the \texttt{DataIter} and 
\texttt{BatchSampler}
so that they can accept a user-specified batch size at runtime and generate a corresponded sample batch.

\vspace{.4em}
\noindent\textbf{Measuring Worker Execution Status.}
To implement LB-BSP, at the worker side, we need to obtain the batch processing time $t$ and state information (e.g., CPU or memory usage). Acquiring $t$ is easy in \emph{dynamic-graph} based frameworks like PyTorch, but is challenging for \emph{static-graph} based frameworks like TensorFlow or MXNet. In TensorFlow, each iteration (forward or backward propagation and synchronization) is executed as a whole with \texttt{tf.Session()}, which cannot be decomposed with simple instructions.
We choose to profile the batch processing time from the \texttt{Timeline} log once each iteration finishes. Instead of respectively measuring the communication time and computation time which might overlap with each other, we directly calculate batch processing time as the \emph{iteration time} minus the \emph{synchronization waiting time} (i.e., duration of the \texttt{sync\_token\_q\_Dequeue} operation).
Besides, to measure the CPU or memory usage we resort to the Python \texttt{psutil} \cite{psutil} library, and to measure the GPU memory usage we adopt the \texttt{tf.contrib.memory\_stats.BytesInUse()} operation.

\vspace{.4em}
\noindent\textbf{Thrift RPC Protocol.} The \texttt{BatchSizeManager} can be located in a dedicated machine or co-located with the PS process on an existing server of the cluster.
For efficient communication between the \texttt{BatchSizeManager} and workers, we employ Apache Thrift \cite{thrift}, a lightweight RPC protocol developed by Facebook, and we create a thread pool to serve the worker requests in parallel.

\vspace{.4em}
\noindent\textbf{Online NARX Training.}
The NARX models we use for CPU clusters are written in Keras~\cite{Keras}, a high-level neural network API.
Accurate NARX training requires collecting enough samples, so we enable NARX prediction only after the first $500$ iterations. In practice we find that $500$ samples are enough for accurately training our NARX models, which are quite simple (\cref{sec:narx}). Within the first $500$ iterations, we can use EMA or, if that training job is recurring, the past NARX models trained in former runs. 
For fast convergence, we initialize NARX models by \emph{model reusing}~\cite{yang2017deep}---with the models trained even for other workers or ML jobs. 
Besides, we also adopt \emph{early stopping}~\cite{yao2007early} and in practice we found that most training processes terminate within 10 steps.

\section{Evaluation}
\label{sec:evaluation}
\begin{figure*}
	\parbox[t]{16.5cm}{
		\centering
		\subfloat[Hardware Efficiency]
		{
			\label{fig:macro_perupdate}
			\includegraphics[width=0.29\textwidth]{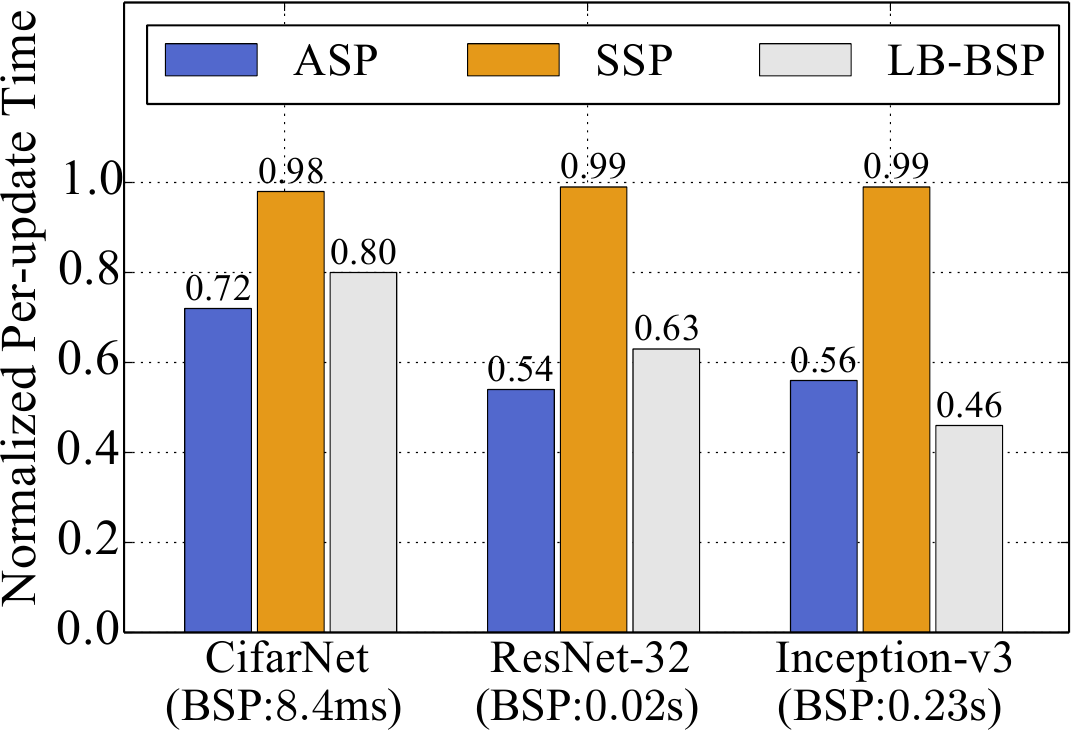}
			\vspace{-0.1in}
		}\hfill
		\subfloat[Statistical Efficiency]
		{	
			\label{fig:macro_numofupdates}
			\includegraphics[width=0.3\textwidth]{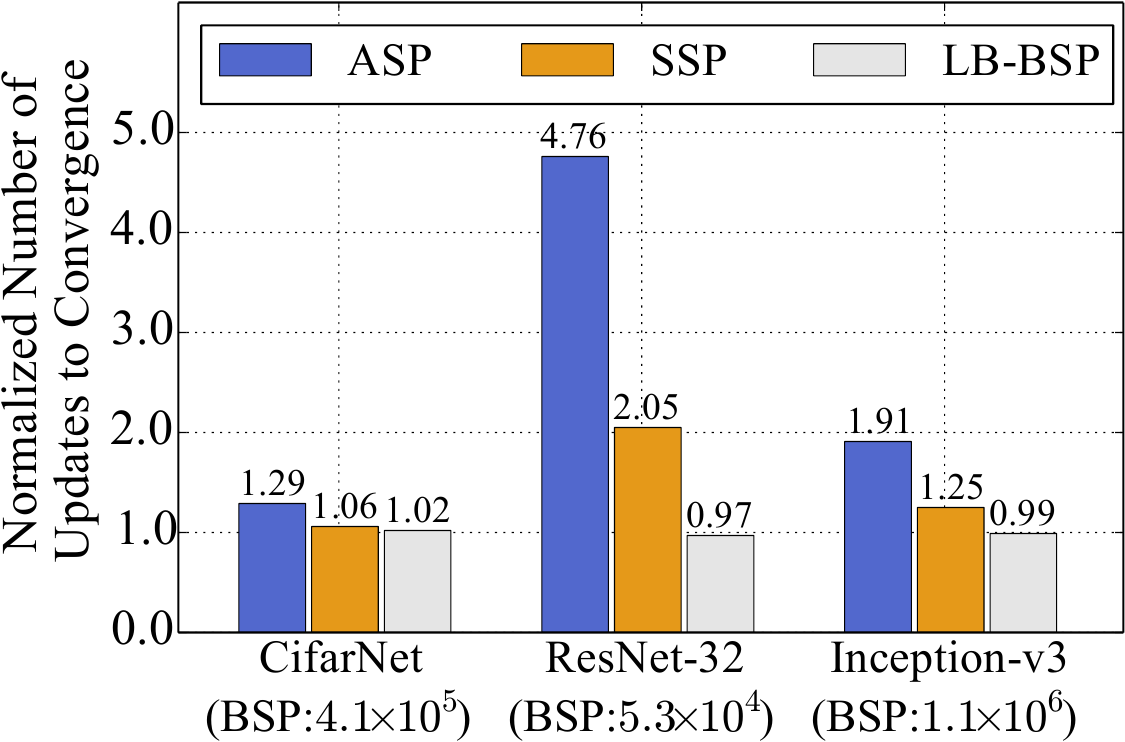}
			\vspace{-0.1in}
		}\hfill
		\subfloat[Overall Convergence Efficiency]
		{	
			\label{fig:macro_runtime}
			\includegraphics[width=0.29\textwidth]{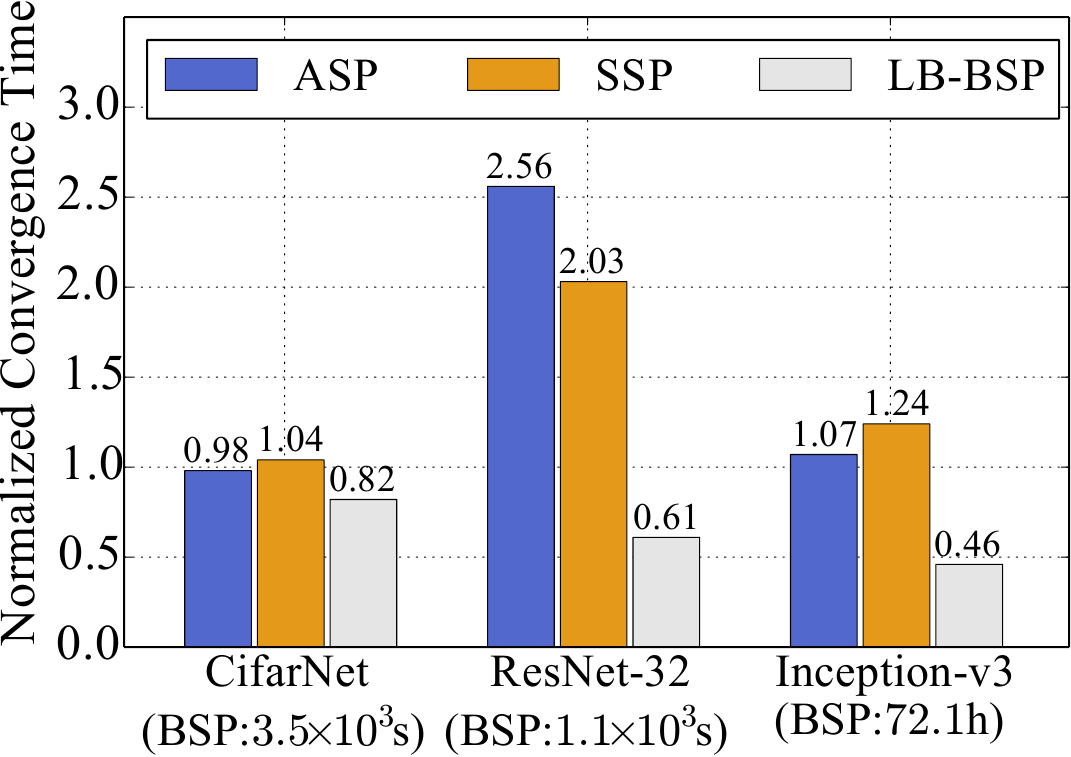}
			\vspace{-0.1in}
		}
		\vspace{-0.15in}
		\caption{Training efficiency under different worker coordination schemes in Cluster-A.}
		\label{fig:macro}
		\vspace{-0.05in}
	}
\end{figure*}

In this section we systematically evaluate the performance of LB-BSP in non-dedicated clusters. 
We start with the end-to-end (\cref{sec:eval_gpu}) and micro-benchmark (\cref{sec:micro_gpu}) evaluations in GPU clusters. Then we resort to model training with leftover resources in production CPU clusters (\cref{sec:eval_production}), with a deep dive analysis of the NARX prediction approach (\cref{sec:deepdive}). Finally we evaluate the overhead of our LB-BSP solution in \cref{sec:eval_overhead}.

\subsection{End-to-End Result in GPU Cluster}
\label{sec:eval_gpu}
\noindent\textbf{Experimental Setup.} As elaborated in \cref{sec:background}, there do exist some scenarios (e.g., due to budget limitation or fairness policy) where model training has to be conducted in heterogeneous GPU clusters. To emulate such scenarios, we build \textbf{Cluster-A}, a heterogeneous GPU cluster with 16 Amazon EC2 instances: four \texttt{p3.2xlarge} instances (each with one Tesla V100 GPU), four \texttt{g3.4xlarge} instances (each with one Tesla M60 GPU), four \texttt{p2.xlarge} instances (each with one Tesla K80 GPU), and four \texttt{g2.2xlarge} instances (each with one GRID K520 GPU). On each instance we run a worker process and a collocated PS shard under TensorFlow 1.4.0.

With Cluster-A, we train the CifarNet~\cite{krizhevsky2009learning} and ResNet-32~\cite{he2016deep} model on CIFAR-10 dataset, and Inception-V3~\cite{szegedy2016rethinking} model on ImageNet dataset~\cite{ILSVRC15} (containing 1.28 million of training images of 1000 classes). For simplicity each worker locally hosts a full dataset copy.
The initial batch size of each worker is set to $128$ for CIFAR-10, and $32$ for ImageNet. The initial learning rate is set to $0.01$. 
The schemes evaluated in this part are BSP, ASP, SSP\footnote{
	By default, TensorFlow does not support SSP. 
	We implemented it with a \texttt{worker-coordinator} module over the Thrift RPC protocol, which enforces fast workers to wait if the slowest one is 5 iterations behind.} 
and LB-BSP, and we defer the comparisons with FlexRR and redundant execution to CPU clusters\footnote{ 
	We exclude FlexRR from GPU evaluations because it is not compatible with the tensor-based processing style of GPUs, and redundant execution is also excluded because its behavior in heterogeneous GPU clusters is trivial---always ignoring the most inferior GPU worker(s).}.
We measure the overall training efficiency as well as the hardware and statistical efficiency. The results are summarized in Fig.~\ref{fig:macro}, where BSP is the baseline and all the values displayed are normalized by that under BSP.

\vspace{.4em}
\noindent\textbf{Hardware Efficiency.} The metric we use for hardware efficiency is \emph{per-update time}---the average time it takes for the PS to receive one gradient update. For BSP and LB-BSP, per-update time is the average iteration time divided by the number of workers. Fig.~\ref{fig:macro_perupdate} shows that
LB-BSP remarkably outperforms BSP and SSP, and this is consistent with our analysis in \cref{sec:sync_schemes}. Interestingly, LB-BSP is even $15\%$ better than ASP in hardware efficiency when training Inception-V3 ---we will explain that with micro-benchmark evaluations in \cref{sec:micro_gpu}.

\begin{figure}[t]
	\vspace{-0.15in}
	\centering
	\subfloat
	{
		\includegraphics[width=0.222\textwidth]{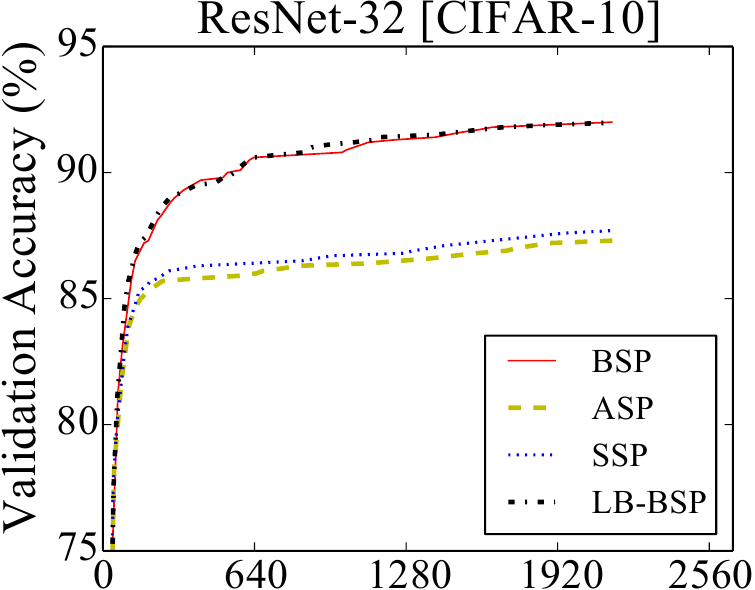}
	}\hfill
	\subfloat
	{	
		\includegraphics[width=0.225\textwidth]{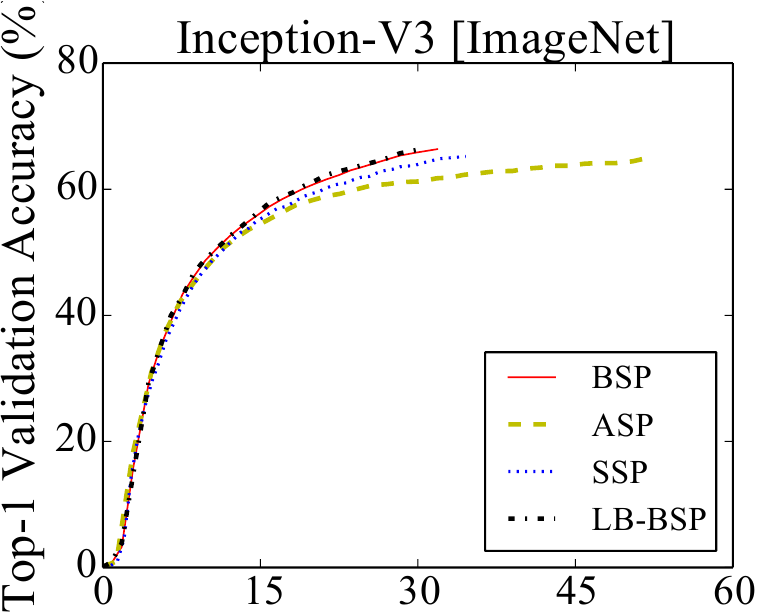}
	}
	\vspace{-0.15in}
	\caption{Test accuracy against training epochs.}
	\label{fig:accuracy_curve}
	\vspace{-0.3cm}
\end{figure}

\begin{figure}[t]
	\vspace{-0.05in}
	\centering
	\subfloat
	{
		\includegraphics[width=0.222\textwidth]{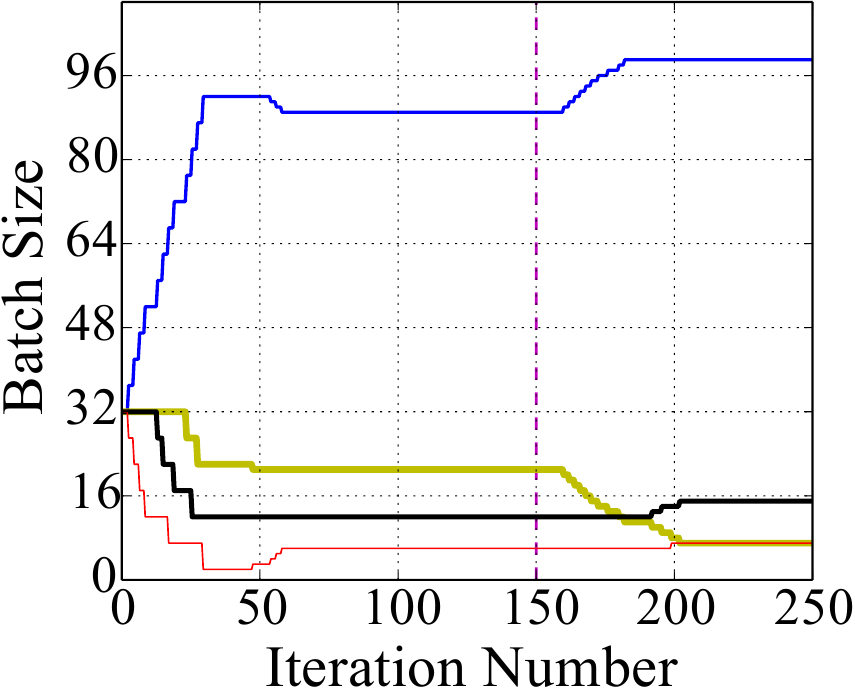}
	}\hfill
	\subfloat
	{	
		\includegraphics[width=0.225\textwidth]{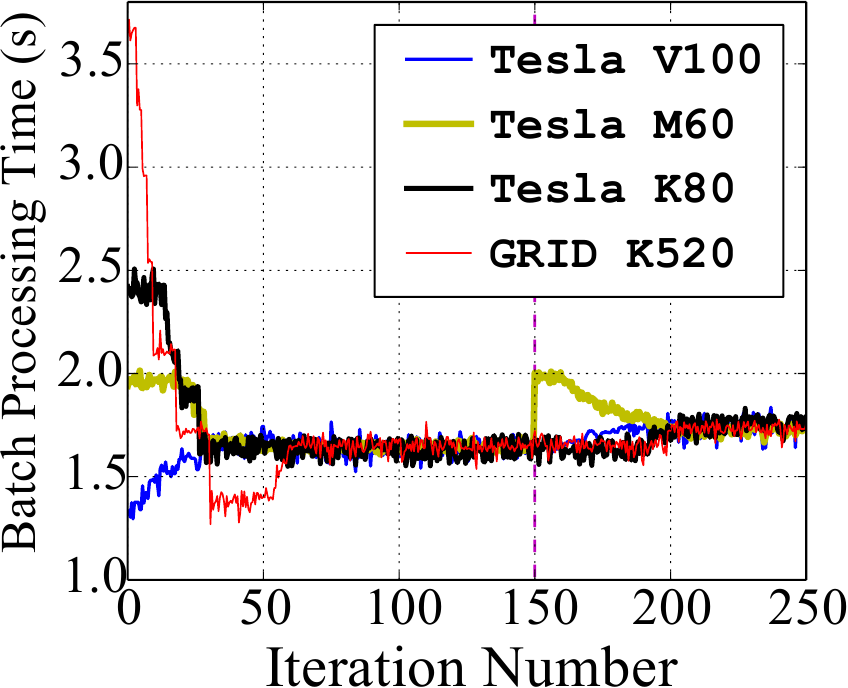}
	}
	\vspace{-0.05in}
	\caption{Instantaneous batch size and batch processing time of the four heterogeneous GPU workers under LB-BSP. At iteration $30$, Alg.~\ref{alg:gpu-lbbsp} enters the fine-tune phase, and batch sizes of the four workers stabilize at $(89,21,12,6)$. Later at iteration 150, bandwidth of the M60 GPU worker is reduced to emulate a migration-caused locality degradation, and the four workers then enter a new equilibrium.}
	\label{fig:lbbsp_bs_curve}
	\vspace{-0.2cm}
\end{figure}

\vspace{.4em}
\noindent\textbf{Statistical Efficiency.} Statistical efficiency is measured as the number of updates required to reach the target accuracy. We set different near-optimal accuracy targets for different models: $0.80$ for CifarNet, $0.86$ for ResNet-32, and $0.65$ for Inception-V3. As shown in Fig.~\ref{fig:macro_numofupdates}, for each model under LB-BSP, the number of updates required to reach the target accuracy is almost identical with that under BSP. In contrast, ASP and SSP require up to $4.76\times$ and $2.05\times$ the number of BSP to make that accuracy. 

Moreover, ASP and SSP fall behind not only in the convergence speed, but also in the the ultimate accuracy attained.
Fig.~\ref{fig:accuracy_curve} shows the convergence curves of ResNet-32 and Inception-V3, where an epoch is a full pass of all the samples in the CIFAR-10 or ImageNet dataset. For ResNet-32, we find that the accuracy made by ASP or SSP gets stuck below $0.88$, while BSP and LB-BSP successfully make the ideal accuracy of $0.92$. Also, for Inception-V3, BSP and LB-BSP already surpass ASP and SSP even for reaching a sub-optimal accuracy target $0.65$\footnote{Inception-V3 is not trained to the ideal accuracy ($78.8\%$) due to our budget limitation. For reference an existing work~\cite{chen2016revisiting} has confirmed that BSP can yield a higher final accuracy than ASP even in homogeneous clusters.}.

\begin{figure*}
	\parbox[t]{5.6cm}{\null
		\centering
		\hspace{-0.35cm}
		\begin{footnotesize}
			\begin{tabular}{@{}ccc@{}}\toprule
				\begin{tabular}{@{}c@{}}\textbf{Instance} \\ \textbf{Type} \end{tabular} & \begin{tabular}{@{}c@{}}\textbf{CPU, Mem} \\ \textbf{(core, GiB)} \end{tabular} &\textbf{Num} \\ \midrule
				\texttt{m4.2xlarge} &(8, 32) & 17 \\ \hdashline
				\texttt{c5.2xlarge} &(8, 16) & 10 \\ \hdashline
				\texttt{r4.2xlarge} & (8, 61) &2 \\ \hdashline
				\texttt{m4.4xlarge} &(16, 64) & 2 \\ \hdashline
				\texttt{m4.xlarge} & (4, 16) &1 \\ 
				\bottomrule
				\vspace{-0.02in}
			\end{tabular}
		\end{footnotesize}
		\captionof{table}{Cluster-B composition.}
		\label{tbl:composition}
	}
	\parbox[t]{5.3cm}{
		\vspace{-0.15in}
		\centering
		\subfloat
		{
			\hspace{-0.4cm}
			\label{fig:cifar10_converge_curve}
			\includegraphics[width=0.32\textwidth]{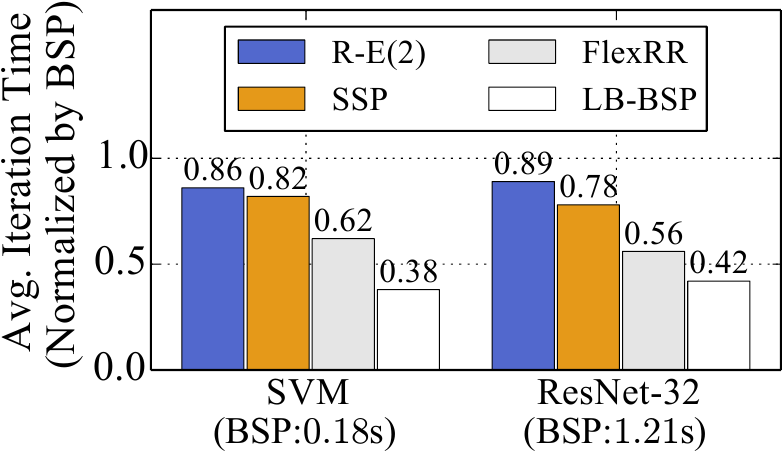}
		}
		\vspace{-0.3cm}
		\caption{Iteration time with different schemes in Cluster-B.}
		\label{fig:macro_cpu}
	}
	\hfill
	\parbox[t]{5cm}{\null
		\vspace{-0.1in}
		\hspace{-0.6cm}
		\centering
		\includegraphics[width=0.31\textwidth]{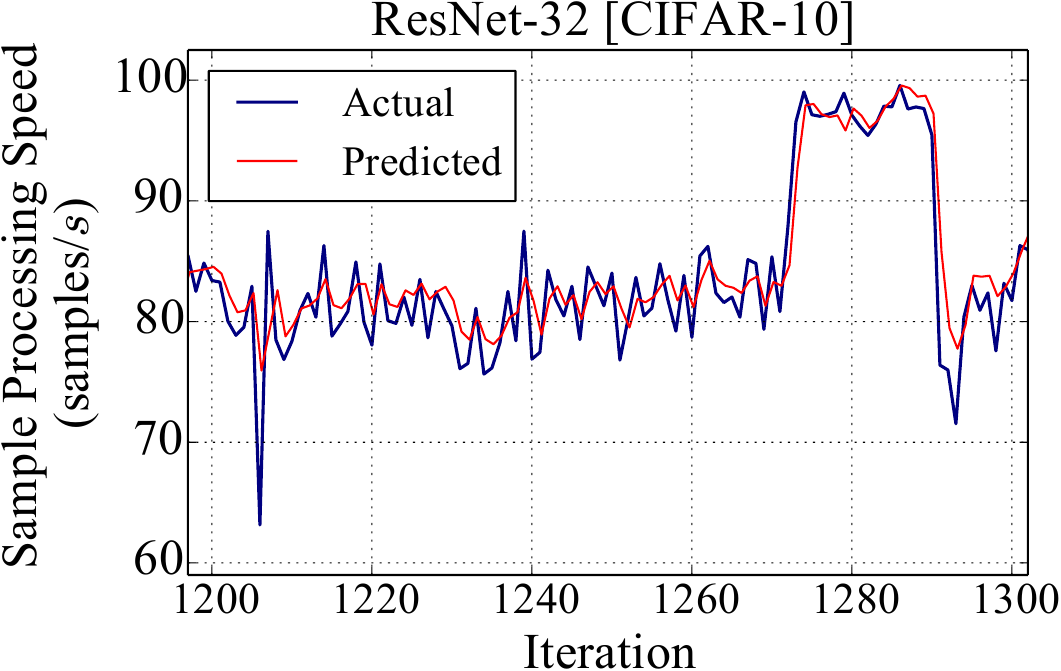}
		\hspace{0.2cm}
		\vspace{-0.3in}
		\caption{NARX prediction result on an \texttt{m4.2xlarge} instance.}%
		\label{fig:prediction_curve}
	}
	\vspace{-0.1in}
\end{figure*}

\vspace{.2em}
\noindent\textbf{Overall Convergence Efficiency.} Fig.~\ref{fig:macro_runtime} shows the overall time required to reach the target accuracy, where LB-BSP surpasses the second best by up to $54\%$. Therefore, it's highly rewarding to employ LB-BSP in such heterogeneous GPU clusters. 
We further train ResNet-32 in a 12-node GPU cluster without the \texttt{p3.2xlarge} instances, and the iteration speedup of LB-BSP over BSP reduces (from $37\%$ in Fig.~\ref{fig:macro_perupdate}) to $28\%$. Thus, the more heterogeneous the cluster is, the more necessary it is to adopt LB-BSP for load balancing.

\subsection{Micro-benchmark in GPU Cluster}
\label{sec:micro_gpu}

To further understand LB-BSP behavior from the micro level, we scale down Cluster-A to contain only 4 GPU instances each of a distinct type, and measure the instantaneous worker batch size and batch processing time when training Inception-V3. As shown in Fig.~\ref{fig:lbbsp_bs_curve}, LB-BSP gradually increases the batch size of the most powerful worker (i.e., the one with the Tesla V100 GPU), with the batch sizes of the other workers correspondingly decreased.
Finally an equilibrium is reached where all the workers share almost the same batch processing time. 
Note that after iteration $30$, the LB-BSP algorithm (Alg.~\ref{alg:gpu-lbbsp}) enters the fine-tune phase, where the batch sizes of the four workers gradually stabilize at $(89,21,12,6)$.

Fig.~\ref{fig:lbbsp_bs_curve} also helps to elaborate why LB-BSP can even outperform ASP in hardware efficiency. By yielding 26 samples (from 32 to 6), the worker with K520 GPU has its batch processing time reduced by nearly 2s; in contrast, the worker with V100 GPU, after incorporating as many as 57 samples, only suffers an increase of 0.5s. Therefore, by allowing advanced workers to process more samples and achieve a higher utilization with negligible slowdown, LB-BSP can reduce the average cost to process one sample and improve the hardware efficiency\footnote{
	This also holds for CifarNet and ResNet-32 in Fig.~\ref{fig:macro_perupdate}, but their iterations are shorter and the GPU random perturbations are more significant, rendering ASP still better than LB-BSP in hardware efficiency.}.

Furthermore, we evaluate LB-BSP applicability in shared GPU clusters, and emulate the network variation caused by locality degradation when conducting cross-machine(rack) worker migration.
In Fig.~\ref{fig:lbbsp_bs_curve}, at iteration 150 we bound the bandwidth of the worker with M60 GPU to 2.5Gbps (from EC2-provisioned 10Gbps, with the \texttt{wondershaper}~\cite{wondershaper} tool). The LB-BSP algorithm learns that shortly and then gradually adjusts workers' batch size to reach another equilibrium. This confirms that LB-BSP can work well in multi-tenant GPU clusters with job migrations from time to time.

\vspace{-0.08in}
\subsection{Evaluation in Shared CPU Cluster}
\label{sec:eval_production}
\vspace{-0.02in}

\vspace{.4em}
\noindent\textbf{Experimental Setup.}
As elaborated in \cref{sec:cpu}, non-neural-network or not-so-urgent models may be trained in non-dedicated CPU clusters. To evaluate LB-BSP performance in such scenarios, we manually created \textbf{Cluster-B}, a heterogeneous CPU cluster that emulates the shared production environment where ML models are trained with the dynamic leftover resources ~\cite{reiss2012heterogeneity,cortez2017resource,li2018ease,hazelwood2018applied}. Cluster-B is built based on a one-hour snapshot of Google Trace~\cite{reiss2012heterogeneity}. That trace 
discloses the machine configurations (CPU/Memory capacities in \emph{normalized} form) of a production cluster from Google, together with the information of all the involved jobs/tasks during a selected month---including their resource consumptions and start/end times. 

More specifically, we \emph{scale down} the totally 12,583 machines to 32 EC2 instances, with the former's \emph{hardware heterogeneity proportionally} preserved---by accordingly selecting the instance types and the quantity of each type, as summarized in Table~\ref{tbl:composition}.
Meanwhile, \emph{resource dynamicity} of that Google cluster is also emulated: we randomly map each instance to a machine in the Google cluster,
and launch a set of faked tasks sharing identical behaviors (i.e., start/end times \& CPU/memory consumptions) with those submitted to that Google machine. 

Regarding the models, we train SVM on a malicious URL dataset~\cite{ma2009identifying}, and ResNet-32 on CIFAR-10 dataset. The batch size for SVM training is $10\%$ of the whole URL dataset (as in \cite{jiang2017heterogeneity}), and is 128 for ResNet-32. 

The schemes evaluated are Redundant Execution (R-E), SSP, LB-BSP and FlexRR. The first three schemes are implemented in TensorFlow. For R-E, we add two \texttt{c5.2xlarge} instances (also in resource contention with some faked tasks) to Cluster-B as the backup worker, and only collect $32$ gradients returned the earliest in each iteration. R-E is supported in TensorFlow with the \texttt{replicas\_to\_aggregate} parameter. In SSP, the bound of iteration gap is still set to $5$, as in \cref{sec:eval_gpu}. 
Regarding FlexRR, since it is not open-sourced and its sequential sample processing manner is not supported in current ML frameworks~(\cref{sec:load-balancing}), we choose to emulate it in PyTorch. We generate tiny batches each containing only one sample, and then encapsulate them into logical batches of designated size; such a logical batch can then be processed in a sequential manner. Meanwhile, each worker's helper group is set to be all the remaining workers, and for the other setups (e.g., progress check frequency, trigger condition of load reassignment), we follow the suggested values in \cite{harlap2016addressing}.

\vspace{.5em}
\noindent\textbf{Hardware Efficiency.} 
Fig.~\ref{fig:macro_cpu} shows the average iteration time when training SVM and ResNet-32 under different schemes in Cluster-B. For fair comparison, each value in Fig.~\ref{fig:macro_cpu} is normalized by the BSP performance in the corresponded framework (sequential-processing-style PyTorch for FlexRR, and TensorFlow for the others). 
As in Fig.~\ref{fig:macro_cpu}, R-E and SSP speed up the training iterations only marginally, because they focus on either worst-case or transient stragglers, but the stragglers in Cluster-B span a wide range of degrees and durations.
Meanwhile, regarding FlexRR, its performance is better but not the best,
because its decentralized load balancing decisions are not optimal, and meanwhile it suffers non-negligible measurement, negotiation and load reassignment overheads (\cref{sec:load-balancing}). 
In contrast, LB-BSP can bring a speedup of $62\%$ for SVM and $58\%$ for ResNet-32 over BSP, outperforming the second best (FlexRR) by up to 38.7\%. 
Thus, given that LB-BSP can also make the optimal statistical efficiency (\cref{sec:eval_gpu}), it can surely lead to the best convergence efficiency among all the schemes evaluated. 

\vspace{-0.05in}
\subsection{NARX Performance Deep Dive}
\label{sec:deepdive}
\vspace{-0.02in}

\begin{table}[t]
	\centering
	\renewcommand{\arraystretch}{1}
	\footnotesize
	\begin{footnotesize}
		\begin{tabular}{@{}lccc@{}}\toprule
			\textbf{Method} & \textbf{Configuration} & \textbf{RMSE} & \begin{tabular}{@{}c@{}}\textbf{Avg. Iteration Time} \\ \textbf{(Normalized by BSP)} \end{tabular}  \\ \midrule
			\textbf{Memoryless} & - & 11.85 & $0.58$ \\ \hdashline
			\textbf{EMA} & $\alpha$=0.2 & 7.85 & $0.48$ \\ \hdashline
			\textbf{ARIMA} & ($p$,$d$,$q$)=(2,2,1) & 9.67 & 0.52 \\ \hdashline
			\textbf{SimpleRNN} & look-back=2 & 8.34 & 0.49 \\ \hdashline
			\textbf{LSTM} & look-back=2 & 9.19 & 0.51 \\ \hdashline
			\textbf{NARX} & look-back=2 & 4.78 & 0.42 \\
			\bottomrule
		\end{tabular}
	\end{footnotesize}
	\caption{RMSE and iteration speedup when applying different prediction approaches in LB-BSP.}
	\label{tbl:prediction_performance}
	\vspace{-0.2in}
\end{table}
In this part, we further evaluate the prediction performance of the NARX model (\cref{sec:narx}) we use in Cluster-B.

\vspace{.2em}
\noindent\textbf{Visual Analysis.} To get a \emph{visual} understanding of the NARX prediction performance, we randomly select a period (iteration 1200$\sim$1300) from the ResNet-32 training process on one \texttt{m4.2xlarge} instance of Cluster-B; the \emph{actual} and \emph{predicted} sample processing speeds are presented in Fig.~\ref{fig:prediction_curve}. From it we observe that the benefit of NARX is twofold. 
On the one hand, NARX is robust to non-deterministic \emph{transient} perturbations: when there are ``\emph{spikes}'' (like the sharp wave around iteration 1206) in the actual speed curve, the predicted curve fluctuates much less. This is because NARX predicts also with the worker's available memory and CPU amounts, which are relatively stable during those ``spikes''. 
On the other hand, when the actual speed increases not for randomness but for non-transient deterministic factors like increased CPU or memory resources (e.g., around iteration 1270), the predicted speed can promptly catch up.

\vspace{.5em}
\noindent\textbf{Comparison with Other Approaches.} We further compare NARX with other approaches surveyed in \cref{sec:narx}, as listed in Table~\ref{tbl:prediction_performance}. The \emph{memoryless} method means to take last iteration's sample processing speed as the predicted one. Regarding the EMA approach, the smoothing factor $\alpha$ (weight of the latest observation) is set to be $0.2$. As for the statistical prediction approach---ARIMA, its \emph{order of the autoregressive model} ($p$), \emph{degree of differencing} ($d$), and \emph{order of the moving average} ($q$) are respectively set to 2, 2 and 1, based on the model selection techniques~\cite{ozaki1977order}. Finally, for SimpleRNN (plain RNN) and LSTM, their \emph{look-back} window size is set to 2, the same as in NARX. 

Then, we replace the NARX approach with those candidate prediction approaches, and re-train ResNet-32 model under LB-BSP in Cluster-B. For each approach, we record the average \emph{root-mean-square error} (RMSE) of the prediction results and the corresponded iteration time (\emph{normalized} by the iteration time under BSP). 
From Table~\ref{tbl:prediction_performance},
the NARX approach, with the ability to perceive CPU/memory resource variations, attains the best performance---it surpasses the second best by around $40\%$ in RMSE and $15\%$ in average iteration time. 

\vspace{-0.05in}
\subsection{System Overhead and Scalability}
\label{sec:eval_overhead}

In TensorFlow, LB-BSP introduces two extra procedures: first to extract batch processing time from execution logs (e.g., the \texttt{Timeline} object), and second to conduct RPC communication between the \texttt{BatchSizeManager} and workers. Yet, they won't slow down GPU workers because of our non-blocking design~(\cref{sec:implementation}), and here we measure the slowdown caused by those two extra procedures in CPU clusters.

\begin{figure}[t]
	\centering
	\vspace{-0.15in}
	\subfloat
	{
		\includegraphics[width=0.22\textwidth]{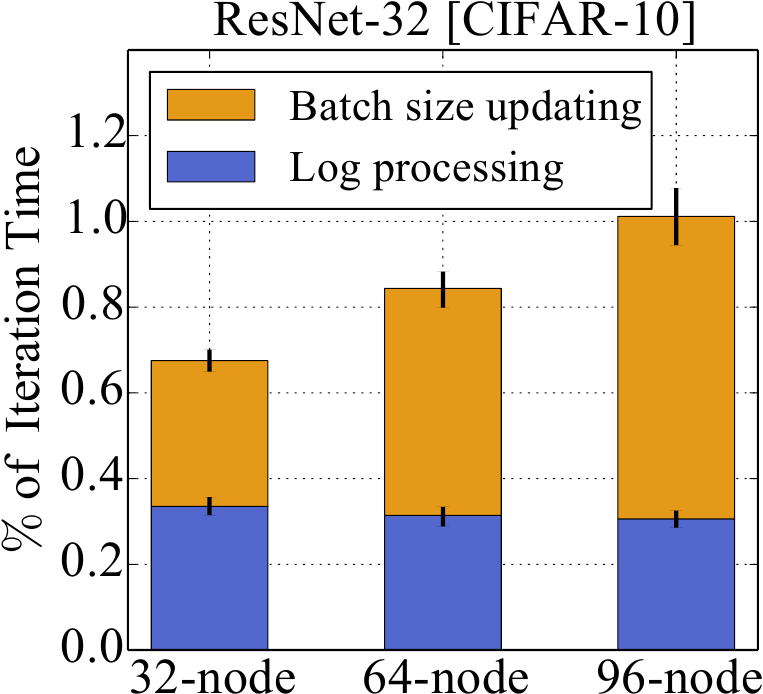}
	}\hfill
	\subfloat
	{	
		\includegraphics[width=0.22\textwidth]{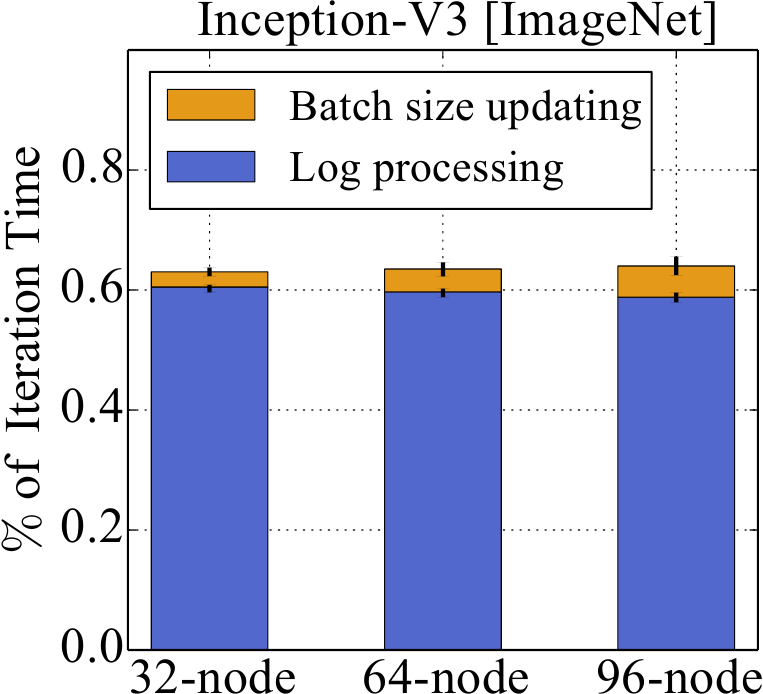}
	}
	\vspace{-0.05in}
	\caption{LB-BSP overheads in CPU clusters.}
	\label{fig:overhead}
	\vspace{-0.2in}
\end{figure}

We respectively train ResNet-32 and Inception-V3 model for 1000 iterations in three CPU clusters---a \emph{homogeneous} cluster with 33 \texttt{c5.2xlarge} instances (32 worker nodes and 1 separate node hosting both the PS and \texttt{BatchSizeManager}), and then its \emph{enlarged} version with a \emph{doubled/tripled} number of workers. Fig.~\ref{fig:overhead} shows the average time respectively spent on log processing and batch size updating, \emph{normalized} by the iteration time (error bars show the $5^{th}$/$95^{th}$ percentile). Even in the largest cluster with 96 workers, the total overheads are less than 1.1\% of the iteration time for both models. This indirectly confirms that, performance of the PS is almost not affected by the co-located \texttt{BatchSizeManger}.

\section{Additional Related Work}
\label{sec:related}

Besides the related work in \cref{sec:sync_schemes}, in this part we discuss some additional related work on \emph{batching}. Batching is necessary when processing long-lasting streaming inputs or training models with large datasets. For big data streaming systems \cite{zaharia2013discretized,toshniwal2014storm}, 
some works \cite{das2014adaptive,zhang2016adaptive} have explored how to adaptively adjust the batching interval when faced with dynamic data rates or operating conditions.
For iterative model training, some \cite{devarakonda2017adabatch,de2017automated} have proposed to adaptively increase batch size during the training process to yield faster convergence.
Yet, those works don't involve workload allocation among parallel workers, and are thus orthogonal to LB-BSP.
\section{Conclusion}
\label{sec:conclusion}
In this work, we propose LB-BSP to load-balance distributed model training workloads in a semi-dynamic manner, by speculatively apportioning the load on the workers according to their temporal processing capability. LB-BSP is tailor-made respectively for both CPU and GPU clusters, and our experiments on Amazon EC2 have shown clear evidence that it can effectively eliminate stragglers in non-dedicated clusters, speeding up model convergence by over $50\%$.
\section*{Acknowledgment}
\label{sec:acknowledgment}

The research was supported in part by RGC GRF grants under the contracts
16206417, 16207818 and 26213818.  Qizhen Weng was supported in part by the
Hong Kong PhD Fellowship Scheme.


\bibliographystyle{ACM-Reference-Format}
\bibliography{main}

\end{document}